\documentclass[journal]{IEEEtran}
\ifCLASSINFOpdf
\else
\fi
\usepackage{amsmath, amsthm, amssymb}
\usepackage{algpseudocode}
\usepackage{algorithm}
\usepackage{multirow}
\usepackage{array}
\usepackage[lofdepth,lotdepth]{subfig}
\usepackage{booktabs}
\usepackage{bm}
\usepackage{graphicx}
\usepackage{xcolor}
\usepackage{soul}
\usepackage[utf8]{inputenc}
\usepackage[figuresright]{rotating}
\usepackage{subfig}
\usepackage{float}
\usepackage{subfloat}
\usepackage{stfloats}
\usepackage{cases}
\usepackage{mdwmath}

\usepackage{tcolorbox}
\usepackage{authblk}

\usepackage[normalem]{ulem}
\usepackage{siunitx}
\usepackage{hyperref}

\begin{document}
 
\newtheorem{theorem}{Theorem}
\newtheorem{lem}{Lemma}
\newtheorem{cor}{Corollary}

\theoremstyle{definition}
\newtheorem{defn}[theorem]{Definition} 
%
\title{Convolutional Sparse Support Estimator Network (CSEN)\\ \textit{From energy efficient support estimation to learning-aided Compressive Sensing}}
\author[1]{Mehmet Yama\c{c}}
\author[1]{Mete Ahishali}
\author[2]{Serkan Kiranyaz}
\author[1]{Moncef Gabbouj}
\affil[1]{Tampere University, Faculty of Information Technology and Communication Sciences, Tampere, Finland}
\affil[2]{Department of Electrical Engineering, Qatar University, Qatar}

\maketitle

\begin{abstract}
Support estimation (SE) of a sparse signal refers to finding the location indices of the non-zero elements in a sparse representation. Most of the traditional approaches dealing with SE problem are iterative algorithms based on greedy methods or optimization techniques. Indeed, a vast majority of them use sparse signal recovery techniques to obtain support sets instead of directly mapping the non-zero locations from denser measurements (e.g., Compressively Sensed Measurements). This study proposes a novel approach for learning such a mapping from a training set. To accomplish this objective, the Convolutional Support Estimator Networks (CSENs), each with a compact configuration, are designed. The proposed CSEN can be a crucial tool for the following scenarios: (i) Real-time and low-cost support estimation can be applied in any mobile and low-power edge device for anomaly localization, simultaneous face recognition, etc. (ii) CSEN's output can directly be used as "prior information" which improves the performance of sparse signal recovery algorithms. The results over the benchmark datasets show that state-of-the-art performance levels can be achieved by the proposed approach with a significantly reduced computational complexity.
\end{abstract}


%
\IEEEpeerreviewmaketitle

\section{Introduction}
\label{Introduction}
%
%
%
%
\IEEEPARstart{S}{parse} Representation or Sparse Coding (SC) denotes representing a signal as a linear combination of only a small subset of a pre-defined set of waveforms. Compressive Sensing (CS) \cite{CS1,CS2} can be seen as a special form of SC while a signal, $\mathbf{s} \in \mathbb{R}^{ d}$ which has a sparse representation, $\mathbf{x}  \in  \mathbb{R}^{n}$ in a dictionary or basis $ \mathbf{\Phi} \in \mathbb{R}^{ d\times n}$, can be acquired in a compressed manner using a linear dimensional reductional matrix, $\mathbf{A}  \in \mathbb{R}^{m \times d}$. Therefore, this signal can also be represented in a sparse manner in the dictionary, $ \mathbf{D}  \in \mathbb{R}^{m \times n}$, (that can be called equivalent dictionary \cite{equivalent}, where $m << n$, and typically assumed to be full-row rank), which is the matrix multiplication of the measurement matrix, $\mathbf{A}$ and pre-defined dictionary, $\mathbf{\Phi}$, i.e., $\mathbf{D} = \mathbf{A} \mathbf{\Phi}$. In SC literature, \textit{signal synthesis} refers to producing a signal, $ \mathbf{y}=\mathbf{Dx}  \in \mathbb{R}^m$, using a sparse code, $\mathbf{x} \in \mathbb{R}^n$ and a pre-specified dictionary, $ \mathbf{D}$. On the other hand, \textit{signal analysis} deals with finding the sparse codes, $ \mathbf{x} $ from the given measurements, $\mathbf{y}$, with respect to the dictionary $\mathbf{D}$ \cite{Elad}. Sparse Support Estimation or simply \textit{Support Estimation} (SE)  \cite{SE1,SE2,SE3}, refers to finding the location indices of non-zero elements in SCs. 
In other words, it is the localization of the smallest subset of the atoms, which are the basis waveforms in the dictionary, whose linear combination represents the given signal well enough. On the other hand, \textit{sparse Signal Recovery} (SR) refers to finding the values of these non-zero elements of SCs. SE and SR are intimately linked in such a way that the SE of a sparse signal is first performed, then an SR will be trivial using the ordinary Least Squares optimization. Actually, this is the main principle of most greedy algorithms \cite{OMP,cosamp}

\begin{figure*}[t]
 \centering
  \includegraphics[width=0.95\linewidth]{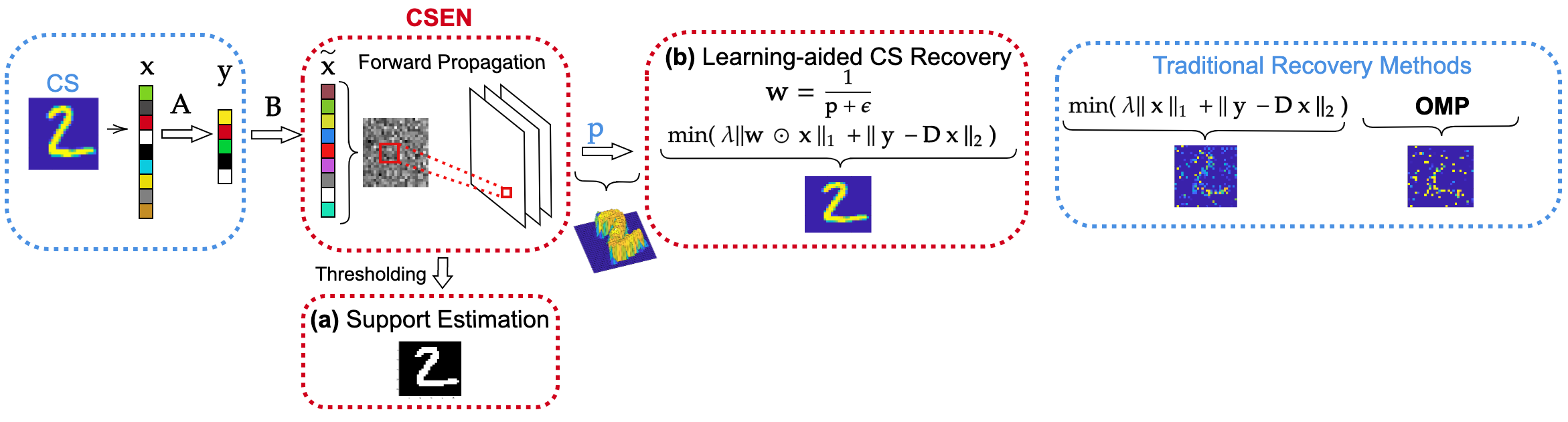}
 \caption{The proposed CSEN with two potential applications: a) (bottom-left) Sparse Support Estimation b) (top-middle) Learned aided CS-sparse signal reconstruction with CSEN vs.  (top-right) traditional recovery methods; i) OMP \cite{OMP} and ii) $\ell_1$-minimization. }
\label{fig:CSEN general}
\end{figure*}

The literature that purely targets SE is relatively short compared to extensive studies on sparse signal recovery \cite{Volkan}. Many existing works, first apply a coarse SR using existing SR methods, and then SE can be easily performed if SE is the main objective. Indeed, there are many applications where computing the support set is more important than computing the magnitudes of SCs. For instance, in an SR based classification (SRC) \cite{SRC2}, such as face recognition \cite{SRC1}, the training samples are stacked in the dictionary in such a way that a subset of the columns consists of the samples of a specific class. As another example, in cognitive radio systems, only a small ratio of all spectrum is occupied for a given time interval. Therefore, finding the occupied spectrum (i.e., the support set) is the primary concern \cite{SS1,SS2}. Similarly, in a ground-penetrating radar imaging system, finding the location of the target is more important than predicting the actual signal magnitudes \cite{radar1}.

In this study, a novel Convolutional Support Estimator Network (CSEN) is proposed with two primary objectives as in Figure \ref{fig:CSEN general}. First, this approach enables learning-based non-iterative support estimation with minimal computational complexity. To accomplish this, we use two compact Convolutional Neural Network (CNN) configurations, both of which are designed without the dense layers \cite{fully}. The proposed CSENs are trained to optimize the support estimations. To our knowledge, this is the first study that proposes a learning-based approach for non-iterative support estimation.  Hence, in order to perform comparative evaluations we train the following state-of-the-art CS signal reconstruction deep neural networks as the support estimators: 1) ReconNet \cite{reconnet} that originally works on the spatial domain and, 2) the Learned AMP (LAMP) \cite{Lamp}, that is the deep version of AMP \cite{AMP}, which is the state-of-the-art optimization scheme working on sparse domain. An extensive set of experiments over three benchmark datasets has demonstrated that the proposed CSEN approach outperforms both deep counterparts, especially dealing with a structural sparse signal. In the first experimental setup, we simulate a CS system making data acquisition from the MNIST data set in different measurement rates.  Moreover, the proposed SE system is shown to improve the SE performance when compared to its deep counterparts, especially in low measurement rates and imperfect sparsity (in the case of CS of approximate sparse signal or noisy environment). Furthermore, CSEN is tested on a well-known support recovery problem, where face recognition is performed based on sparse codes \cite{SRC2}. We use two benchmark datasets, Yale-B \cite{yale}, and CelebA \cite{CelebA}, in our experiments. Comparative evaluations performed against the two state-of-the-art dictionary-based (representation-based) face recognition methods in the literature, SR based face recognition \cite{SRC2}, and collaborative learning \cite{collaborative}, have demonstrated that the proposed CSEN  approach outperformed both methods. \\ \indent As for the second objective, we focus on an alternative usage of CSENs. Instead of using them as support estimators, which naturally requires the hard-thresholding of the network outputs, these outputs can be directly used as prior information about the sparse signals. It is a well-known fact that having prior information about the non-zero locations such as the probability map, $p(x)$ (or simply $\mathbf{p}$), on the support set, could improve the conventional signal recovery algorithms \cite{priori1}. However, in many cases, it is not clear how to obtain such prior information in advance. The most common usage of such a system appears in dynamical sparse recovery \cite{recursive1}, where previous support estimations can be used as priors for the next estimation. In this study, we have demonstrated that CSEN outputs can be a better alternative for the prior information of the non-zero locations. Therefore, CSEN is now used as a learning-aided CS reconstruction scheme, where the prior information comes directly from the CSEN outputs. A wide range of experiments shows that this approach has great potential to improve SR performance of traditional approaches for sparse signal recovery problems. As mentioned above, we used CS imaging simulation, but this time signal reconstruction error is compared with state-of-the-art conventional SR approaches. Figure \ref{fig:CSEN general} illustrates a representative graph of two different applications of CSENs; a) performing SE from CS measurement vector, $\mathbf{y}$, and b) the output of CSEN is used as the side information, $\mathbf{p}$, which gives the estimated probability of being non-zero for each index. In this simple illustration we assume that the hand-writing signal '$2$' is sparse in spatial domain such that $\mathbf{\Phi} = \mathbf{I}$; therefore, $\mathbf{D} = \mathbf{A} \mathbf{I} = \mathbf{A}$ and $\mathbf{ B }$ is a denoiser matrix such as $\mathbf{D}^T$, or  $\left ( \mathbf{D}^T \mathbf{D} + \lambda \mathbf{I} \right )^{-1} \mathbf{D}^T$ where $\lambda$ is the regularization parameter.

The rest of the paper is organized as follows. In Section \ref{Notations}, we start by giving mathematical notation that is used in this article. A brief overview of sparse representation and compressive sensing theory, with emphasis on state-of-art sparse signal recovery and support estimation techniques, will be given in Section \ref{Related Work}. In the same section, we also introduce case studies of support estimation that are chosen for this work. Then we discuss the limitations of existing support estimator techniques. In Section \ref{Proposed}, we will present the proposed learned based SE scheme and the two compact CSEN models. Experimental evaluations of the study will also be given at the end of this section, which we can divide into three main categories according to the case studies: (i) Basic support estimation performance evaluation on MNIST dataset that is performed to compare CSENs with the aforementioned state-of-art deep networks. (ii) SE based face recognition performance evolution of proposed SE with an emphasis on how CSEN based SE has the ability to improve the classical representation-based approaches. (iii) Performance comparison of classical compressing sensing reconstruction techniques and proposed learned-aided SR in terms of both speed and reconstruction accuracy. Having theoretical and experimental analysis, in Section \ref{Discussion} we will present a more detailed discussion on how the proposed scheme differs from the state-of-the-art SR and SE techniques, pros and cons, and possible usage scenarios with an emphasis on the flexibility of proposed CSEN in different scenarios. Finally, the conclusions are drawn in Section~\ref{conclusion}.

\section{Notations}
\label{Notations}
In this work, we define the $\ell_p$ norm of any vector $\mathbf{x} \in \mathbb{R}^n$ as $\left \| \mathbf{x} \right \|_{\ell_p^n} =   \left (  \sum_{i=1}^n \left \vert x_i \right \vert^p \right )^{1/p}$ for $p \geq 1$. The $\ell_0$-norm of the vector $\mathbf{x} \in \mathbb{R}^n$ is given as $\left \| \mathbf{x} \right \|_{\ell_0^n} = \lim_{p \to 0} \sum_{i=1}^n \left \vert x_i \right \vert^p = \# \{ j: x_j \neq 0 \}$ and the $\ell_{\infty}$ is defined as 
$\left \| \mathbf{x} \right \|_{\ell_{\infty}^n} =  \max_{i=1,...,n} \left (   \left | x_i \right | \right )$. A signal $\mathbf{s}$ can be defined as a strictly $k$-sparse signal if it can be represented with less than $k+1$ non-zero coefficients in a proper basis $\mathbf{\Phi}$, i.e., $\left \|  \mathbf{x} \right \|_0 \leq k$, where $ \mathbf{s}= \mathbf{ \Phi}\  \mathbf{x}$. We also define a sparse support set or simply support set, $\Lambda \subset \{1,2,3,...,n \} $, as the set of indices that represents the non-zero coefficients, i.e., $\Lambda := \left \{ i:  x_i =0 \right \}$. The complement of support set, $\Lambda$, with respect to $\{1,2,3,...,n \}$ is given as $\Lambda^c = \{1,2,3,...,n \} \setminus \Lambda$. In this manner, $\mathbf{x}_{\Lambda}  \in \mathbb{R}^{  \left | \Lambda \right |   }   $  is a vector consisting of non-zero elements of $\mathbf{x} \in \mathbb{R}^n$, where  $\left | \Lambda \right |$ refers to the number of the non-zero coefficients . Similarly, $\mathbf{M}_{\Lambda}  \in \mathbb{R}^{ m \times  \left | \Lambda \right |   }   $ denotes a matrix that consists of the columns of a matrix $\mathbf{M}  \in \mathbb{R}^{ m \times  n   }   $ indexed by support ${\Lambda}$.

\section{Related Work}
\label{Related Work}
CS theory claims that a signal $\mathbf{s}$ can be sensed using far fewer linear measurements $m$ than Nyquist/Shannon based traditional methods use, $d$, i.e.,
 \begin{equation}
     \mathbf{y} = \mathbf{A} \mathbf{s} = \mathbf{A}\mathbf{ \Phi} \mathbf{x} =\mathbf{ D} \mathbf{x} \label{CS},
 \end{equation}
where $\mathbf{A} \in \mathbb{R}^{m \times d}$ is the measurement matrix and $\mathbf{D} \in \mathbb{R}^{m \times n}$ is called the equivalent dictionary. It can be demonstrated that sparse representation, 
\begin{equation}
\min_\mathbf{x} ~ \left \| \mathbf{x }\right \|_{0}~ \text{subject to}~ \mathbf{D} \mathbf{x} = \mathbf{y} \label{sparse_rep}
\end{equation}
is unique if $m \geq 2k$ \cite{spark} and $\left \| \mathbf{x }\right \|_{0} \leq k$.  In brief, the uniqueness of the sparse representation in Eq. \eqref{sparse_rep} shows that any $k$-sparse signal pair can still be distinguished in the equivalent dictionary, $\mathbf{D}$. However, the problem in Eq. \eqref{sparse_rep} is that this is a non-convex problem and known to be NP-hard. The most common approach is the relaxation of the $\ell_0$ norm to the closest convex norm, which is $\ell_1$-norm,
\begin{equation}
     \min_\mathbf{x} \left \| \mathbf{x} \right \|_1 ~s.t. ~ \mathbf{x} \in \mho \left (\mathbf{ y} \right ) \label{Eq:l1}
\end{equation}
where $\mho \left ( \mathbf{y} \right ) = \left \{ \mathbf{x}: \mathbf{D} \mathbf{x}=\mathbf{y} \right \}$, which is known as Basis Pursuit \cite{BP}. The surprising result of the CS theory is that even if the exact recovery of the signal, $\mathbf{s}$ was not possible by using the minimum norm solution, a tractable solution is possible using \eqref{Eq:l1}, when $\mathbf{D}$ satisfies some properties such as Restricted Isometry Property \cite{candesRIP} and $m > k (\log (n/k))$. 

However, the signal of interest, $\mathbf{x}$, is not perfectly $k$-sparse but approximately sparse in most of the cases. In addition, CS measurements, most probably, are corrupted by an additive noise during data acquisition, quantization, etc.  As a result, we handle $ \mathbf{y} =\mathbf{D} \mathbf{x} + \mathbf{z} $, where $\mathbf{z}$ is additive noise. In this case, the constraint can be relaxed by setting $\mho \left ( \mathbf{y} \right ) = \left \{ \mathbf{x}: \left \|  \mathbf{D}\mathbf{x}-\mathbf{y}  \right \|_2  \leq \epsilon \right \}$ which is known as Basis Pursuit Denoising \cite{BPDN} or Dantzig Selector \cite{dantzig} if we set  $\mho \left (\mathbf{ y} \right ) = \left \{\mathbf{ x}: \left \| \mathbf{D}^T\left ( \mathbf{y}-\mathbf{D}\mathbf{x} \right ) \right \|_{\infty } \leq \lambda   \right \}$. In the noisy case, even exact recovery of sparse signal is not possible, stable recovery is well studied in the literature for Basis Pursuit Denoising \cite{candesRIP2} and Dantzig Selector \cite{Eldar,dantzig2}.
We mean by stable recovery is that a stable solution $\hat{\mathbf{x}}$ obeys $\left \| \mathbf{x }- \hat{\mathbf{x}} \right \| \leq \kappa \left \| \mathbf{z} \right \|$ where the $\kappa$ is small constant. Another related formulation is 
\begin{equation}
\label{lasso}
    \min_\mathbf{x}  \left \{ \left \|  \mathbf{D}\mathbf{x}-\mathbf{y} \right \|_2^2 + \lambda \left \|\mathbf{ x} \right \|_1  \right \}    
\end{equation}
which is known as Lasso \cite{lasso} formulation, which is also known to produce stable solution in noisy case and exact solution in noise free case \cite{lasso-stable}. 


\begin{figure}[]
 \centering
  \includegraphics[width=0.79\linewidth]{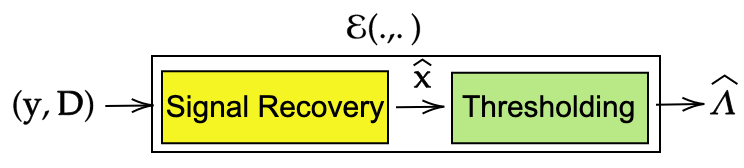}
 \caption{Most common model for a practical support estimator.}
\label{fig:A support Estimator}
\end{figure}

\subsection{Generic Sparse Support Estimation (SE)}
In many application scenarios, detecting the indices of the non-zero coefficients' location, $\Lambda$, is more important than computing these coefficients. To list a few, in a sparse anomaly (either from CS \cite{compressedanomaly} or uniform sampled measurements) detection problem \cite{yamacmalicious}, where a group of users initiates a flooding attack to a communication network (specifically for a VoIP network), detecting the malicious user group (sub-set of all users) is more critical.  Other, CS-based active user detection in the downlink of a CDMA system \cite{CDMA}, and for the uplink of a NOMA \cite{Noma1, Noma2} system. Such systems are believed to play an important role in 5G communication technology. As discussed in Section \ref{Introduction}, other examples may be listed as sparse representation based classifications \cite{SRC2, SRC1} and radar imaging \cite{radar1,radar2}.  

Mathematically speaking, for the linear measurement model given in Eq. \eqref{CS} and with additive noise, $\mathbf{y}= \mathbf{Dx} +\mathbf{z} $, we define the following support estimator $\mathcal{E}(.,.) $,
\begin{equation}
   \hat{\Lambda} =  \mathcal{E}\left (\mathbf{y},\mathbf{D} \right ) 
\end{equation}
where $\hat{\Lambda}$ is the estimated support. For the noise-free case and $\mathbf{x}$ is exactly $k$-sparse, the exact $\Lambda$ recovery performance of an algorithm coincides with the sparse signal recovery performance. This an expected outcome since the unique representation is satisfied when $m >2k$. In the noisy case, even if the exact signal recovery is not possible, it is still possible to recover the support set exactly. In the literature,  several studies have proposed to provide information-theoretical (i.e., the optimal decoder, $\mathcal{E}$'s performance) guarantee conditions for exact  \cite{exact1, SE1, exact4, Volkan},  and partial support estimation \cite{SE3, partial2, Volkan}. However, in most of the practical applications, a tractable signal recovery method is applied first to find an estimation $\hat{\mathbf{x}}$ of the sparse signal $\mathbf{x}$, then a component-wise thresholding is applied to $\hat{\mathbf{x}}$ to compute the estimated support as illustrated in Figure \ref{fig:A support Estimator}.

A common approach is to follow an iterative sparse signal recovery method from the CS literature. For instance, it is proven in \cite{candes2009near} that if $min_{i \in \Lambda} \left | x_i \right | > 8 \sigma \sqrt{2 *log(n)}$, then one can recover the support set exactly using Lasso with $\lambda = 2 \sqrt{2 *log (n)}$, where $\sigma^2$ is variance of the measurement noise. This theorem is valid in the case that the equivalent dictionary satisfies the mutual coherence property defined in \cite{candes2009near}. One may clearly deduce from their results that accurate support estimation is possible via Lasso if the non-zero coefficients' magnitudes are above a certain level determined by the noise. Similarly, the conditions of exact support recovery under noise using OMP are given in \cite{OMP-exact},  and partial support recovery performance bounds of AMP  are in \cite{AMP-partial}.  Along with these SR algorithms in CS literature, which are iterative methods, traditional linear decoders such as   
Maximum Correlation (MC) \cite{MaximumCorrelation}, $\hat{\mathbf{x}}^{MC} = \mathbf{D}^T \mathbf{ y}$ and LMSEE \cite{AMP-partial}, $\hat{\mathbf{x}}^{LMMSE} = \left ( \mathbf{D}^T  \mathbf{D} + \sigma_z^2 \mathbf{I}_{n \times n}  \right )^{-1} \mathbf{D}^T      \mathbf{ y}$  are also used in many applications. The theoretical performance bounds of these methods are also given in \cite{AMP-partial}. 

\subsection{Case study of SE: Representation based classification}
\label{SR classification}
Consider an image from a particular class is queried: It can be expected from the estimated SCs, $\hat{\mathbf{x} }$ to have significant (non-zero) entries which are located in a specific location so that the corresponding columns in the dictionary matrix, $\mathbf{D}$, are the samples from the actual class of the image. This problem is also known as the representation based classification, which is a typical example where the support set location is the main information we are seeking. \\ \indent In \cite{SRC2}, $\ell_1$-minimization is used to obtain such a sparse code to determine the identity of face images. However, in reality such an ideal decomposition is not accomplished in general because face images show a high correlation among different classes. This is why, instead of using the estimated sparse codes, $\hat{\mathbf{x}}$ obtained by an SR technique such as Eq. \eqref{lasso}, the authors propose a four steps solution: (i) Normalization: Normalize all the atoms in $\mathbf{D}$ and $\mathbf{y}$ to have unit $\ell_2$-norm, (ii)  SR: $\hat{\mathbf{x}} = \arg \min_{\mathbf{x}} \left \|\mathbf{ x} \right \|_1 \text{s.t} \left \| \mathbf{y} - \mathbf{D} \mathbf{x} \right \|_2 $, (iii) Residual finding: $\mathbf{e_i} = \left \| \mathbf{y} - \mathbf{D_i}  \mathbf{\hat{x}_i} \right \|_2$, where $\mathbf{\hat{x}_i}$ is the estimated coefficients corresponding the class $i$, (iv)  Class determination: $\text{Class}\left ( \mathbf{y} \right ) = \arg \min \left ( \mathbf{e_i} \right )$. \\ \indent This technique and its similar variants have been reported to perform well not only in face recognition but many other classification problems \cite{human-action, hyperspecral}. Later, the authors of \cite{collaborative} propose to change the second step, from $\ell_1$-minimization to the classical $\ell_2$-minimization; $\mathbf{\hat{x} }= \arg \min_{\mathbf{x}} \left \{ \left \| \mathbf{y }- \mathbf{D}\mathbf{x }\right \|_2^2  + \lambda \left \| \mathbf{x} \right \|_2^2   \right \}$, which has a closed-form solution, $\hat{\mathbf{x}} = \left ( \mathbf{D}^T  \mathbf{D} + \lambda \mathbf{I}_{n \times n}  \right )^{-1} \mathbf{D}^T  \mathbf{ y}$. This collaborative representation based classification (CRC) was reported to achieve a comparable classification performance for different classification problems. For face recognition problems, in particular, the authors reported that high classification accuracies were obtained especially for high measurement rates (MRs).

\begin{figure}[h]
 \centering
  \includegraphics[width=0.79\linewidth]{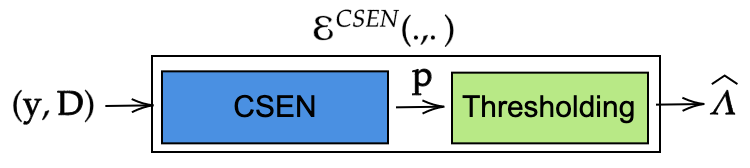}
 \caption{Proposed model for an efficient support estimator.}
\label{fig:proposed Estimator}
\end{figure}
\subsection{Sparse signal reconstruction with side information of support set}
Consider the case where SE is not the main concern but SR is. In case side information is available about the support set, an improvement to $\ell_1$-minimization can be achieved in sparse signal recovery as follows:
\begin{equation}
\label{weighted-lasso}
    \min_\mathbf{x}  \left \{ \left \|  \mathbf{D}\mathbf{x}-\mathbf{y} \right \|_2^2 + \lambda \left \|\mathbf{w \odot x} \right \|_1  \right \}    
\end{equation}
where $\odot$ is element-wise multiplication operator and $\mathbf{w}$ is the predefined cost that imposes the prior information about each element's values. In the concept of modified CS \cite{Modified-CS} and CS with prior information literature, the cost function, $\mathbf{w}$ generally appears in the form of $w_{i} = \frac{1}{p_{i} + \epsilon }$, where $\epsilon > 0$ is a predefined constant and $p_{i}$ is the $i^{th}$ element of the vector $\mathbf{p}$ which is a type of a measure such as prior likelihood \cite{priori1} of the support set, which could represent the probability of $(i)^{th}$ element being non-zero.

\subsection{Limitations of existing Support Estimators}
Both SE and SR algorithms guarantee to perform well if the equivalent dictionary $\mathbf{D}$ satisfies certain properties such as mutual incoherence \cite{coherence}. However, in many practical scenarios, $\mathbf{D}$ fails to satisfy these properties, e.g., in face recognition problem, the atoms of $\mathbf{D}$, vectorized faces, are highly correlated. The second limitation of traditional sparse recovery algorithms is that they are iterative methods and computationally costly. Therefore, the support estimators relying on these sparse recovery algorithms may not be feasible, especially in real-time applications. The third limitation of state-of-the-art SR techniques such as $\ell_1$-minimization is that there is a lower limit for MR (see phase transition \cite{phasetransition}); below this limit, the SR algorithms start to fail completely. This limit generally depends on the wellness of $\mathbf{D}$ (defined by properties such as mutual incoherence \cite{coherence} ). Therefore, SE techniques that build upon an SR algorithm tend to fail if $\mathbf{D}$ does not satisfy the required properties, e.g., if the atoms of $\mathbf{D}$ are highly correlated. 

On the other hand, when it comes to SR techniques leveraging SE as prior information, despite the fact that a good improvement can be achieved using such prior information, most of the works assume that the information is available in advance; however, they do not mention how to obtain such a $\mathbf{p}$.

\section{Convolutional Support Estimator Network}
\label{Proposed}
Recent advance in deep neural networks \cite{LISTA, Lamp} enables a non-iterative solution for the sparse signal recovery. It is often reported that they produce a solution $\mathbf{\hat{x}}$, which is closer to $\mathbf{x}$ than the ones obtained by an iterative approach. They can still work under those measurement rates where classical CS recovery algorithms fail. Nevertheless, a crucial disadvantage is that they require a large number of training samples to achieve a high generalization capability. Second, their complex configuration with millions of parameters causes certain computational complexity issues such as speed and memory problems, especially when they are used in edge devices with limited power, speed and memory.

If one may wish to find only support $\Lambda$ instead of the sign and amplitude of $\mathbf{x}$, a traditional Machine Learning approach would be sufficient. In this study, we propose a support estimator, $\mathcal{E} (.)$, that can be performed by a compact CSEN network. Another crucial objective is to have the ability to learn from a minimal training set with a limited number of labeled data. A typical application where this approach can benefit from is face recognition via sparse representations, where only a few samples of each identity are available.

Let us define a binary mask $\mathbf{v} \in \left \{ 0,1 \right \}^n$, as follows
\begin{subnumcases}
{v_i =}
   1    \hfill & \text{ if $ i \in \Lambda $ } \\
   0 & \text{ else }. 
\end{subnumcases}
Consequently, the problem of finding an estimation $\hat{\mathbf{v}}$ of this binary mask will be equivalent to producing a support estimation $\hat{\Lambda}$, i.e., $\hat{\Lambda} = \left \{  i \in  \left \{  1,2,..,n\right \} : \hat{v}_i =1   \right \}$.

To accomplish this objective, first, the CSEN network with input and output, $\mathcal{P} \left ( \mathbf{y}, \mathbf{D}  \right ): \mathbb{R}^n  \mapsto \left [ 0,1 \right ]^n$, produces a vector $\mathbf{p} $ that gives the information about the probability of each index to be in support set such that $p_i \in \left [ 0,1 \right ]$. Then, the final support estimator, $\mathcal{E} (\mathbf{y}, \mathbf{D} ) $ will produce a support estimation such that $\hat{\Lambda} = \left \{  i \in  \left \{  1,2,..,n\right \} : p_i > \tau   \right \}$, by thresholding $\mathbf{p}$ with $\tau$ where $\tau$ is a fixed threshold.

As shown in Figure \ref{fig:proposed Estimator}, the proposed SE approach is different from the conventional SR based methods, which directly threshold $\hat{\mathbf{x}}$ for support estimation. Moreover, the input-output pair is different. The proposed CSEN learns over $\left ( y^{train}, v^{train} \right )$ to compute $\mathbf{p}$ while the conventional SR methods work with $\left ( y^{train}, x^{train} \right )$ to first make the sparse signal estimation, and then compute support estimation by thresholding it. As evident in Figure \ref{fig:CSEN general}, the application of direct signal recovery may cause noisy estimation of the support codes while the proposed CSEN has the advantage of learning the pattern of the support codes and, therefore, can predict their most-likely location with proper training.

\begin{figure*}[ht!]
 \centering
  \includegraphics[width=0.85\linewidth]{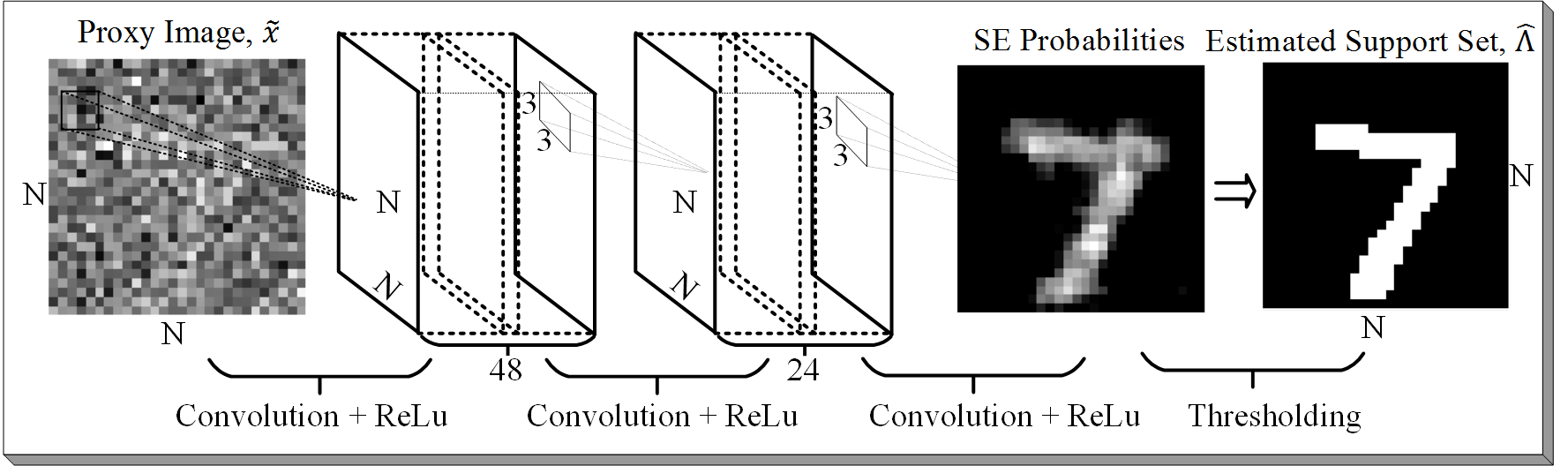}
 \caption{Type-I Convolutional Support Estimator Network (CSEN1).}
\label{fig:compact1}
\end{figure*}

\begin{figure*}[ht!]
 \centering
  \includegraphics[width=0.85\linewidth]{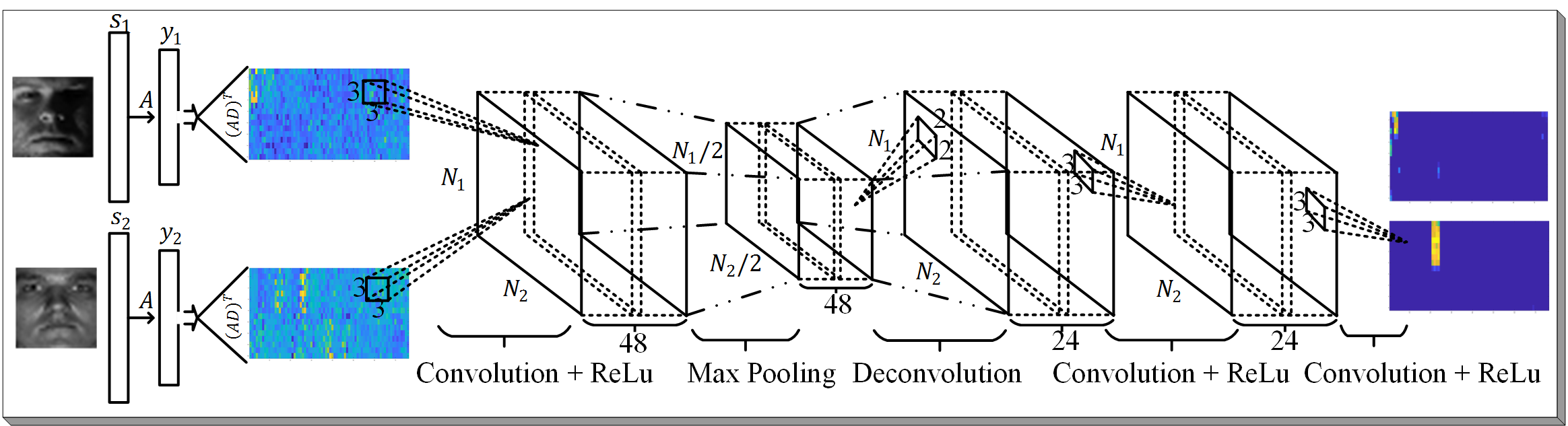}
 \caption{Type-II Convolutional Support Estimator Network (CSEN2).}
\label{fig:compact2}
\end{figure*}

In this study, the proposed CSEN models consist of only convolutional layers in the type of fully convolutional networks \cite{fully} that are trained by optimizing the support estimations. Since the SE problem involves one-to-one mapping, other network types such as Multi-Layer Perceptrons (MLPs) can also be used as in \cite{Lamp}. However, this brings two limitations compared to CSENs: high computational complexity and over-fitting due to the limited training data and number of parameters in the network. In Section \ref{Support Estimation}, it will be shown that such an approach yields a poor generalization and is not robust to noise.

When a CSEN is trained, it learns the following transformation: $ \hat{\mathbf{v}} \leftarrow    \mathcal{P} \left ( \mathbf{\Tilde{x}} \right ) $, where $\hat{\mathbf{v}}$ is the estimation of binary mask representing the estimated support for the signal  $\mathbf{x}$, and the proxy $\mathbf{\Tilde{x}=By}$ with $\mathbf{B=D^T}$, or  $\left ( \mathbf{D}^T \mathbf{D} + \lambda \mathbf{I} \right )^{-1} \mathbf{D}^T$, i.e., the Maximum Correlation and LMMSE formula in \cite{AMP-partial}; and hence, $\mathbf{x}, \mathbf{\Tilde{x}} \in \mathbb{R}^N$. 
First, the proxy $\mathbf{\tilde{x}}$ is reshaped to 2-D plane (e.g., original size of the image or pre-defined search grid). Correspondingly, the proxy $\mathbf{\tilde{x}}$ is convolved with $\mathbf{W_1}$, the weight kernels connecting the input layer to the next layer with $N$ filters to form the input of the next layer with the summation of weight biases $\mathbf{b_1}$ as follows:
\begin{equation}
\mathbf{F_1} = \{S(ReLu(b_1^i + \mathbf{w}_1^i * \Tilde{\mathbf{x}}))\}_{i=1}^{N},
\end{equation}
\noindent where $S(.)$ is the down- or up-sampling operation and $ReLu(x) = max(0, x)$. In more general form, the $k^{th}$ feature map of layer $l$ can be expressed as,
\vspace{-0.15cm}
\begin{equation}
    \mathbf{f_l^k} = \textsc{S}(\textsc{ReLu}(b_l^k + \sum_{i=1}^{N_{l-1}}\textsc{conv2D}(\mathbf{w}_l^{ik}, \mathbf{f}_{l-1}^i, '\textsc{ZeroPad}'))).
\end{equation}
The trainable parameters of the network would be: \\ \noindent $\mathbf{\Theta_{CSEN}}=\big\{ \{\mathbf{w}_1^i, b_1^i\}_{i=1}^{N_1}, \{\mathbf{w}_2^i, b_2^i\}_{i=1}^{N_2}, ... \{\mathbf{w}_L^i, b_L^i\}_{i=1}^{N_L}\big\}$ for a \textit{L} layer CSEN.

In the proposed approach, the Mean-Square Error (MSE) is computed between its binary mask, $\mathbf{v}$, and CSEN's actual output, $ \mathcal{P}_{\Theta}\left (\mathbf{x} \right )_p$ as follows:
\begin{equation}
\label{cost}
E(\mathbf{x}) = \sum_p \left ( \mathcal{P}_{\Theta}\left (\mathbf{x} \right )_p- v_p \right ) ^{2}
\end{equation}
where $v_p$ is the $p^{th}$ pixel of $\mathbf{v}$. The CSEN network is trained using samples in the train data, $D_{train} = \big\{ (\Tilde{\mathbf{x}}^{(1)}, \mathbf{v}^{(1)}), (\Tilde{\mathbf{x}}^{(2)}, \mathbf{v}^{(2)}), ... (\Tilde{\mathbf{x}}^{(s)}, \mathbf{v}^{(s)})\big\}$. Please note that, even if we use MSE as the loss function in the original CSEN design, depending on the application, any other regularization function (e.g., $\ell_1$-norm, mixed norm, etc.) can be added to this cost function. As an example, we present a strategy to approximate the loss function which is group $\ell_1$-norm in addition to MSE.

\section{Results}
\label{Support Estimation}
In order to evaluate the effect of different network configurations, in this study, we use two different CSEN configurations and perform a comprehensive analysis over each of them. Generally, each convolutional layer has a dimension reduction capability with pooling functions. However, the first proposed network architecture consists of only convolutional layers with ReLu activation functions to preserve the sparse signal (e.g., image) dimensions at the output layer. In this configuration (CSEN1), we propose to use three convolutional layers with $48$ and $24$ hidden neurons and $3\times3$ filter size as given in Figure \ref{fig:compact1}. CSEN2 is a slight modification of CSEN1 configuration, as shown in Figure \ref{fig:compact2} by using up- and down-sampling layers. Although this modification increases the number of parameters, in return, it yields substantial performance improvement over MNIST. While the SE performance analysis over MNIST has done using CSEN1 and CSEN2, only CSEN1 results are reported since CSEN2 produces similar recognition rates ($\sim$ 0.001 difference) for face recognition. In any case, both network configurations are compact compared to the deep CNNs that have been proposed recently. For example, the study in \cite{reconnet} proposes ReconNet for SR, which consists of six convolutional layers with $32$ neurons or more in each layer.

Since there is no competing method for SE that is similar to the proposed method, we use the ReconNet \cite{reconnet} in this study on the SE problem by directly giving $\mathbf{\Tilde{x}}$ as the input, and removing the denoiser block at the end for comparative evaluations. Finally, we apply thresholding over the output of ReconNet to generate SE i.e., $\hat{\Lambda}_{\text{R}} = \left \{  i \in  \left \{  1,2,..,n\right \} : \mathcal{P}_{\text{R}} \left ( \mathbf{ \tilde{x}} \right ) > \tau   \right \}$, where $\mathcal{P}_{\text{R}} (.)$ is ReconNet with fully convolutional layers.  
ReconNet is originally a CS recovery algorithm working directly on spatial domain, i.e., $ \mathbf{ \hat{s} } \leftarrow    \mathcal{P} \left ( \mathbf{y} \right ) $ instead of solving them in the sparsifying dictionary, i.e., $\mathbf{\hat{s}} =\mathbf{\Phi} \mathbf{ \hat{x} } $ where $ \mathbf{ \hat{x} } \leftarrow    \mathcal{P} \left ( \mathbf{y} \right ) $. Therefore, ReconNet serves as a deep CSEN approach against which the performance of the two compact CSENs will be compared. Moreover, we also train the state-of-the-art deep SR solution, LAMP, in order to use it over SE problem. For LAMP method, it is possible to predefine the number of layers in advance. For a fair comparison, we have tested the algorithm for three different setups: $2$, $3$, and $4$ layers design using their provided implementation. Next, in the experiments of face recognition based on SR, we consider both speed and recognition accuracy of the algorithms as it is performed only for $\ell_1$-minimization toolbox in \cite{fast}. Thus, in order to perform comparative evaluations, the proposed CSEN approach is evaluated against most of the conventional state-of-the-art SR techniques along with ReconNet. Finally, CSEN2 is applied as a pre-processing step for the CS-recovery to obtain $\mathbf{w}$ in the cost function as illustrated in Figure \ref{fig:CSEN general}.
\begin{figure}[h]
\centering
  \includegraphics[width=0.75\linewidth]{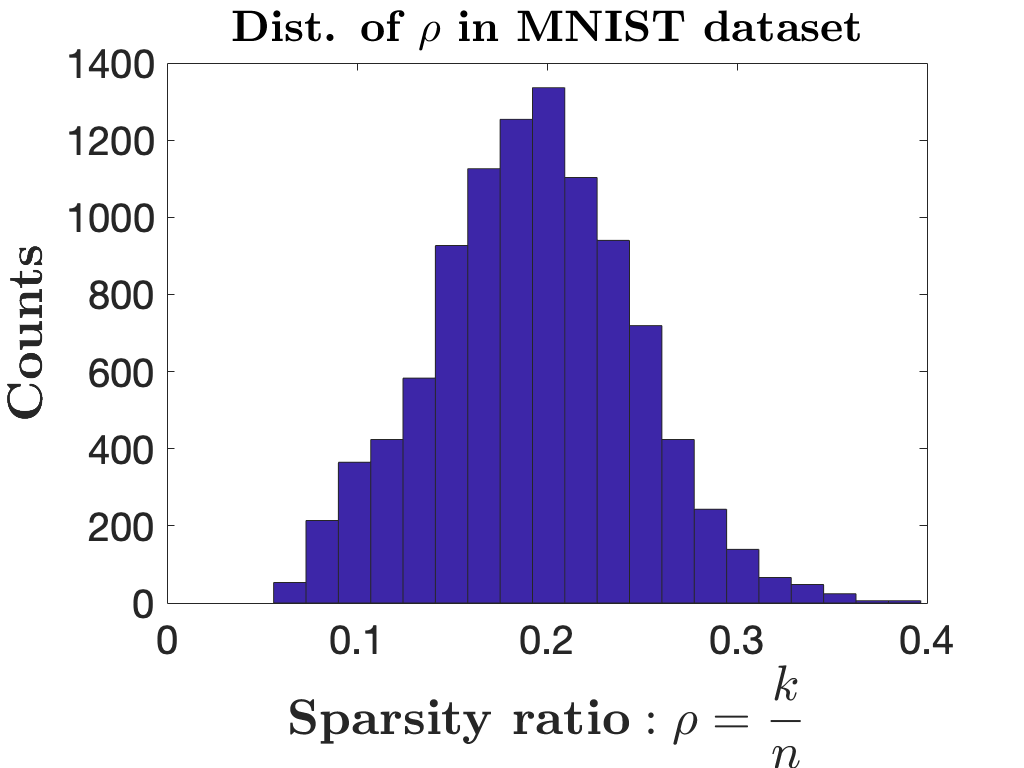}
  \caption{Histogram of $\rho_i$'s obtained from the 10k samples (test set). The vectorized gray scale images, $\mathbf{x}_i$ in MNIST dataset are already sparse in the spatial domain (in canonical basis, i.e., $\Phi =I$) with $\left \| \mathbf{x}_i \right \| \leq k_i$.}
\label{fig:hist_rho}
\end{figure}

The experiments in this study have been carried out on a workstation that has four Nvidia® TITAN-X GPU cards and Intel ® Xeon(R) CPU E5-2637 v4 at 3.50GHz with 128 GB memory. Tensorflow library \cite{abadi2016tensorflow} is used with Python. ADAM optimizer \cite{kingma2014adam} is utilized during the training with the proposed default values of the learning parameters: learning rate, $\alpha=0.001$, and moment updates $\beta_1 = 0.9$, $\beta_2 = 0.999$ with only 100 and 30 Back-Propagation iterations for MNIST and face recognition experiments, respectively.

\subsection{Experiment I: Support Estimation from CS measurements}
For the experiments in this section, MNIST dataset is used. This dataset contains 70000 samples (50K/10K/10K as the sizes of the train/validation/test sets) of the handwritten digits (0 to 9). Each image in the dataset is a $28 \times 28$ pixel resolution with intensity values ranging from 0 (black, background) to 1 (white, foreground). Since the background covers more area than the foreground, each image can be considered as a sparse signal. Mathematically speaking, we may assume that the $i^{th}$ vectorized, $\mathbf{x}_i \in \mathbb{R}^{n=784}$ can be considered as the $k_i$-sparse signal. The sparsity rates of each sample is calculated as $\rho_i = \frac{k_i}{n}$, and its histogram is given in Figure \ref{fig:hist_rho}. We have designed an experimental setup where these sparse signals (sparse in canonical basis) $\mathbf{x}_i$'s are compressively sensed,
\begin{equation}
    \mathbf{y_i} = \mathbf{A} \mathbf{{x_i} }= \mathbf{D} \mathbf{x_i}
\end{equation}
where  $\mathbf{D}=\mathbf{A} \in \mathbb{R}^{m \times n}$ since $\mathbf{\Phi} =\mathbf{ I}$. We calculate the measurement rate as $\mathbf{MR} = \frac{m}{n}$. Therefore, the problem is SE from each CS measurement, i.e., finding $\hat{\Lambda}_i$ from each $\mathbf{y_i}$ in the test dataset.

\begin{figure}[]
\centering
  \includegraphics[width=0.8\linewidth]{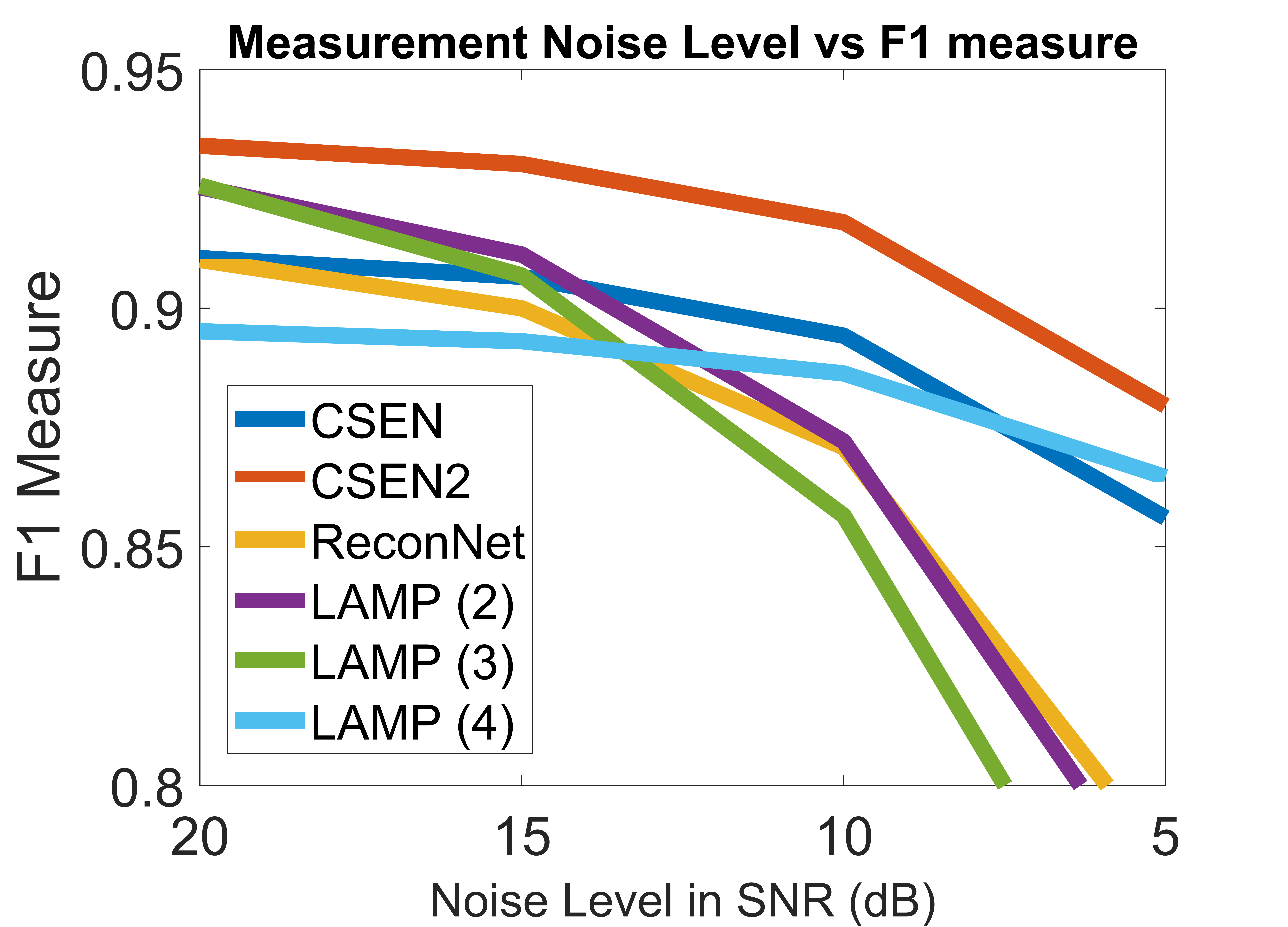}
  \caption{F1 Measure graph of CSEN and Lamp configurations in different noise level at MR = 0.25.}
\label{fig:SNRversusF1}
\end{figure}

\begin{center}
\begin{table}[h] \scriptsize
        \centering
        \setlength\tabcolsep{2.5pt}
        \caption{Support Recovery Performance of Algorithms from the noise-free measurements.}
\bgroup
\def\arraystretch{1.5}
\begin{tabular}{|c|c|c|c|c|c|c|c|c|c|c|c|c|}
\hline
\textbf{MR} & \textbf{0.25} & \textbf{0.1} & \textbf{0.05} & \textbf{0.25} & \textbf{0.1} & \textbf{0.05} & \textbf{0.25} & 
\textbf{0.1} & \textbf{0.05} & \textbf{0.25} & \textbf{0.1} & \textbf{0.05} \\ \hline
            & \multicolumn{3}{c|}{\textcolor[rgb]{0.82,0.01,0.11}{\textbf{F1 Measure}}}              & \multicolumn{3}{c|}{\textcolor[rgb]{0.82,0.01,0.11}{\textbf{Precision}}}               & \multicolumn{3}{c|}{\textcolor[rgb]{0.82,0.01,0.11}{\textbf{Recall}}}                  & \multicolumn{3}{c|}{\textcolor[rgb]{0.82,0.01,0.11}{\textbf{CE}}}                      \\ \hline
\textcolor[rgb]{0.05,0.04,0.89}{\textbf{CSEN}}        & 0.91          & 0.85         & 0.80          & 0.90          & 0.84         & 0.77          & 0.92          & 0.87         & 0.84          & 0.03          & 0.06         & 0.08          \\ \hline
\textcolor[rgb]{0.05,0.04,0.89}{\textbf{CSEN2}}       & \textbf{0.94}          & 0.89         & \textbf{0.84}          & 0.93          & 0.88         & \textbf{0.82}          & \textbf{0.94}          & \textbf{0.90}         & \textbf{0.8}7          & \textbf{0.02}          & \textbf{0.04}         & \textbf{0.06}          \\ \hline
\textcolor[rgb]{0.05,0.04,0.89}{\textbf{ReconNet}}    & 0.90          & 0.85         & 0.79          & 0.89          & 0.82         & 0.76          & 0.90          & 0.87         & 0.83          & 0.05          & 0.06         & 0.09          \\ \hline
\textcolor[rgb]{0.05,0.04,0.89}{\textbf{LAMP (2)}}    & 0.92          & 0.89         & 0.82          & 0.94          & 0.90         & \textbf{0.82}          & 0.89          & 0.87         & 0.83          & 0.05          & 0.05         & 0.08          \\ \hline
\textcolor[rgb]{0.05,0.04,0.89}{\textbf{LAMP (3)}}    & 0.93          & 0.89         & 0.82          & \textbf{0.95}          & 0.90         & \textbf{0.82}          & 0.91          & 0.88         & 0.82          & 0.03          & 0.05         & 0.08          \\ \hline
\textcolor[rgb]{0.05,0.04,0.89}{\textbf{LAMP (4)}}    & 0.93          & \textbf{0.90}         & 0.83          & \textbf{0.95}          & \textbf{0.92}         & \textbf{0.82}          & 0.92          & 0.89         & 0.83          & 0.03          & \textbf{0.04}         & 0.08          \\ \hline
\end{tabular}
\egroup        
\label{tab:performance_noise_free}
\end{table}
\end{center}

For this dataset, the measurement rate, ($\textbf{MR}$) is varied from $0.05$ to $0.25$ in order to investigate the effect of MR on the SE performance. The measurement matrix is then chosen as the "Gaussian", the matrix whose elements $A_{i,j}$ are i.i.d. drawn from $\mathcal{N}\left ( 0,\frac{1}{m} \right )$. It is worth mentioning that the approximate message passing (AMP) algorithm is a well-optimized method for the Gaussian measurement matrix, and LAMP is a learned version of this algorithm. Therefore, they are reported to be state-of-the-art if the measurement matrix is Gaussian but they do not even guarantee the converge for other types of measurement matrices. On the other hand, the comparative performance evaluations against LAMP and deep CS-sparse methods are presented in Table \ref{tab:performance_noise_free}, and the results clearly indicate that the proposed method achieves the best SE performance in terms of F1 measure for MR = 0.25 and 0.05 and comparable for MR = 0.1. The results presented in Table \ref{tab:performance_noise_free} indicate that despite its deep and complex configuration, compact CSENs achieve superior performance levels compared to ReconNet.

\begin{center}
\begin{table}[h] \scriptsize
        \centering
        \setlength\tabcolsep{2.5pt}
        \caption{Support Recovery Performance of Algorithms under 10 $\text{dB}$ measurement noise.}
\bgroup
\def\arraystretch{1.5}
\begin{tabular}{|c|c|c|c|c|c|c|c|c|c|c|c|c|}
\hline
\textbf{MR}       & \textbf{0.25} & \textbf{0.1}  & \textbf{0.05} & \textbf{0.25} & \textbf{0.1}  & \textbf{0.05} & \textbf{0.25} & \textbf{0.1}  & \textbf{0.05} & \textbf{0.25} & \textbf{0.1}  & \textbf{0.05} \\ \hline
\textbf{}         & \multicolumn{3}{c|}{\textcolor[rgb]{0.82,0.01,0.11}{\textbf{F1 Measure}}}      & \multicolumn{3}{c|}{\textcolor[rgb]{0.82,0.01,0.11}{\textbf{Precision}}}       & \multicolumn{3}{c|}{\textcolor[rgb]{0.82,0.01,0.11}{\textbf{Recall}}}          & \multicolumn{3}{c|}{\textcolor[rgb]{0.82,0.01,0.11}{\textbf{CE}}}              \\ \hline
\textcolor[rgb]{0.05,0.04,0.89}{\textbf{CSEN}}     & 0.89          & 0.82          & 0.77          & 0.89          & 0.82          & 0.75          & 0.89          & 0.82          & 0.79          & 0.04          & 0.07          & 0.09          \\ \hline
\textcolor[rgb]{0.05,0.04,0.89}{\textbf{CSEN2}}    & \textbf{0.92} & \textbf{0.86} & \textbf{0.80} & \textbf{0.92} & 0.86          & \textbf{0.80} & \textbf{0.92} & \textbf{0.86} & \textbf{0.82} & \textbf{0.03} & \textbf{0.06} & \textbf{0.08} \\ \hline
\textcolor[rgb]{0.05,0.04,0.89}{\textbf{ReconNet}} & 0.89          & 0.83          & 0.78          & 0.89          & 0.81          & 0.74          & 0.89          & 0.85          & 0.81          & 0.04          & 0.07          & 0.09          \\ \hline
\textcolor[rgb]{0.05,0.04,0.89}{\textbf{LAMP (2)}} & 0.87          & 0.85          & 0.79          & 0.90          & 0.86          & 0.78          & 0.84          & 0.83          & 0.80          & 0.08          & 0.08          & 0.10          \\ \hline
\textcolor[rgb]{0.05,0.04,0.89}{\textbf{LAMP (3)}} & 0.87          & 0.84          & 0.77          & 0.91          & \textbf{0.87} & 0.78          & 0.84          & 0.81          & 0.77          & 0.06          & 0.08          & 0.12          \\ \hline
\textcolor[rgb]{0.05,0.04,0.89}{\textbf{LAMP (4)}} & 0.86          & 0.85          & 0.77          & 0.87          & \textbf{0.87} & 0.78          & 0.85          & 0.82          & 0.77          & 0.08          & 0.07          & 0.12          \\ \hline
\end{tabular}
\egroup        
\label{tab:performance_noisy}
        \end{table}
\end{center}

Furthermore, comparative evaluations are performed when the measurements are exposed to noise in the test set, i.e., $\mathbf{y}_i = \mathbf{D}\mathbf{ x}_i + \mathbf{z}_i$, where $\mathbf{z}_i $ is an additive white Gaussian noise. The results presented in Figure \ref{fig:SNRversusF1} show that SE performances of the LAMP method are adversely affected by increased measurement noise. Their performance gets even worse when the number of layers is increased (i.e., see results for LAMP (2) to LAMP (4)).  CSEN2, on the other hand, achieves the highest F1 measure for all noise levels.

\subsection{Experiment II: Face Recognition based on Sparse Representation}

\begin{figure}[h]
\centering
\includegraphics[width=0.95\linewidth]{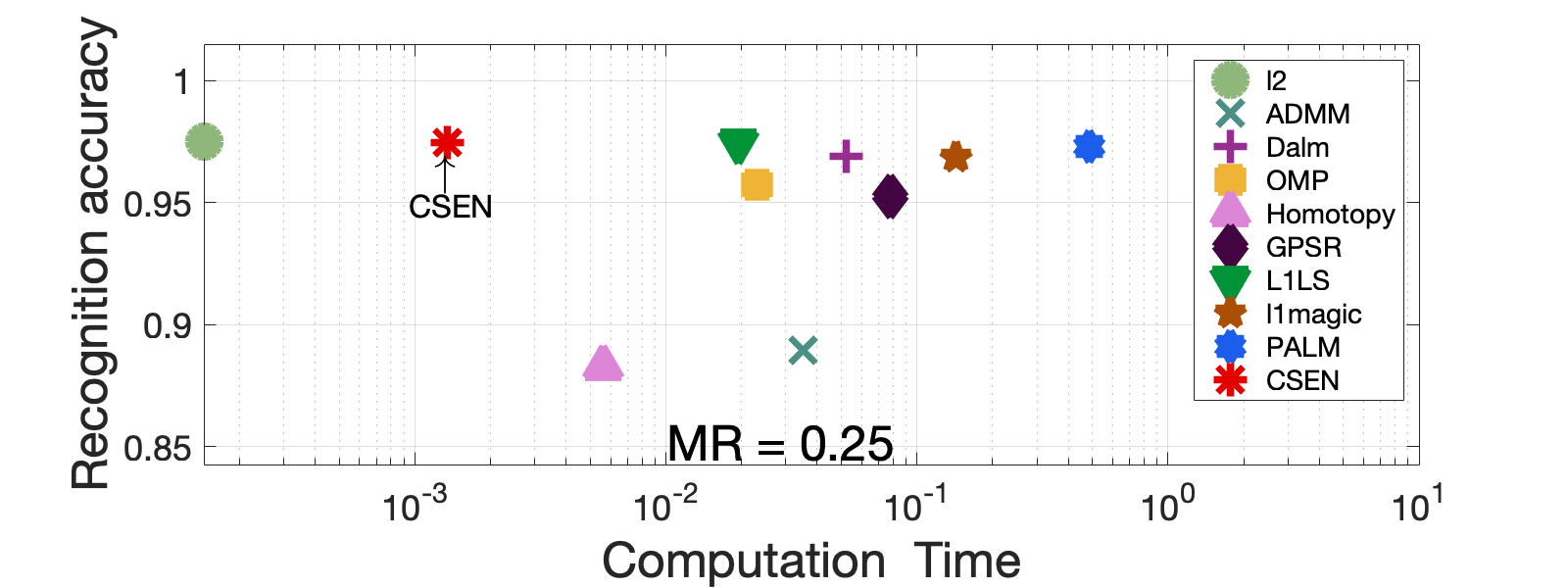}

  \includegraphics[width=0.95\linewidth]{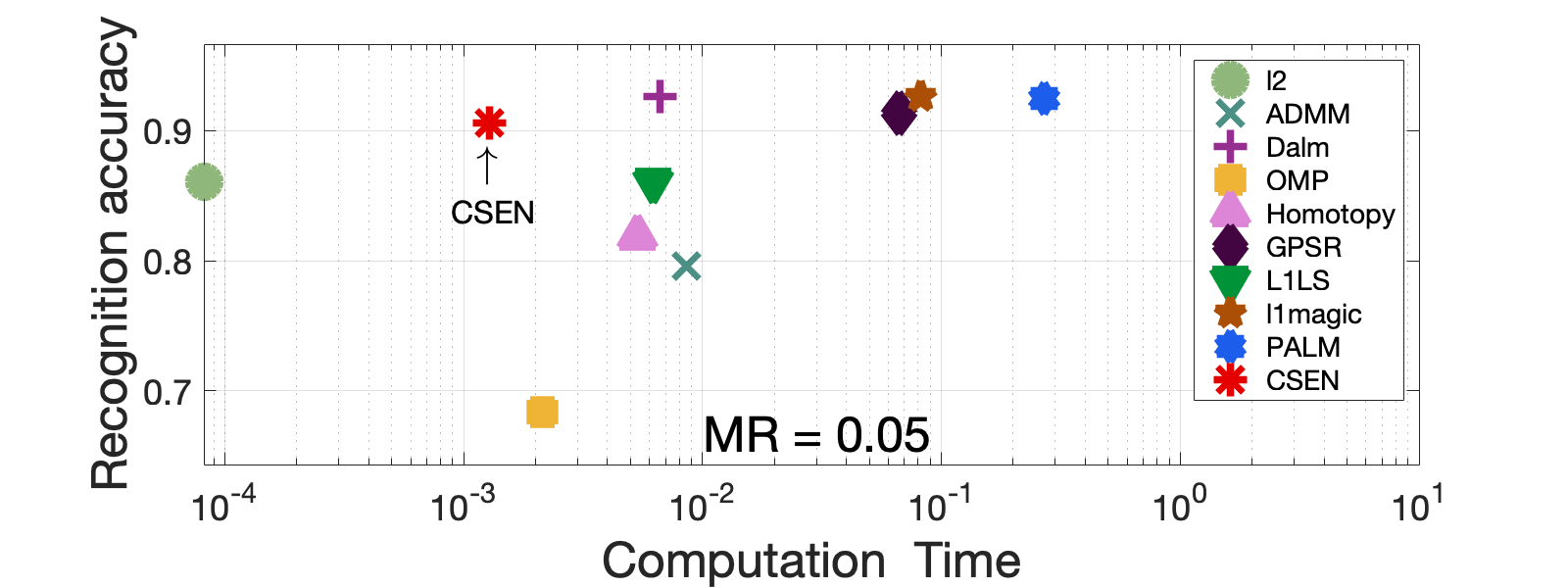}
 
  \includegraphics[width=0.95\linewidth]{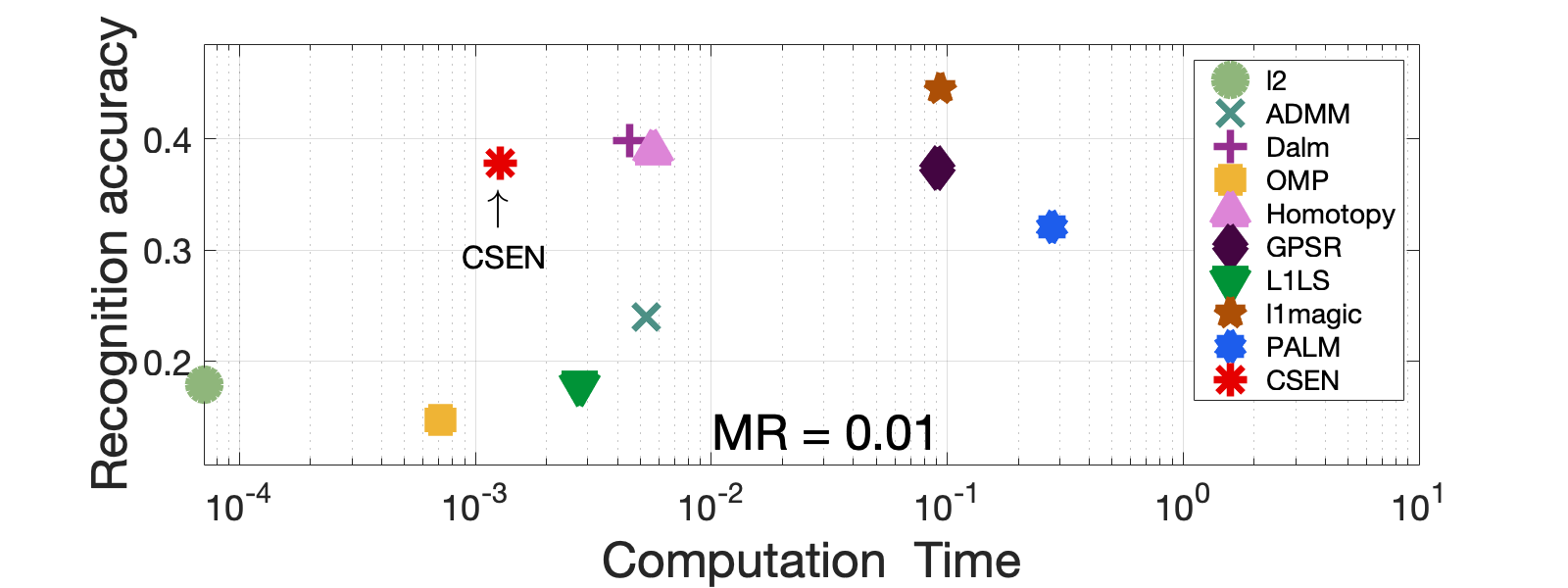}
  
  \caption{Reconstruction accuracy vs. process time comparison of Algorithms in Yale-B database.}
\label{fig:Yale-B}
\end{figure}

As explained in Section \ref{SR classification}, the dictionary-based (representation based) classification could be seen as a SE problem. Therefore, CSEN presents an alternative and better approach to both CRC and SRC solutions. In this manner, the proposed CSEN approach is evaluated against both CRC and the state-of-the-art SRC techniques recently proposed. The algorithms are chosen by considering both their speed and performance on the SR problem, since the speed-accuracy performance of SRC directly depends on the performance of the sparse signal recovery algorithm \cite{fast}, and there is no unique winner to achieve the top performance level for all databases. The proposed method is, of course, not limited to face recognition but can be applied in any other representation based classification problem. For the sake of brevity, in this study, we focus only on the face recognition problem. 

\begin{figure}[h]
\centering
\includegraphics[width=0.95\linewidth]{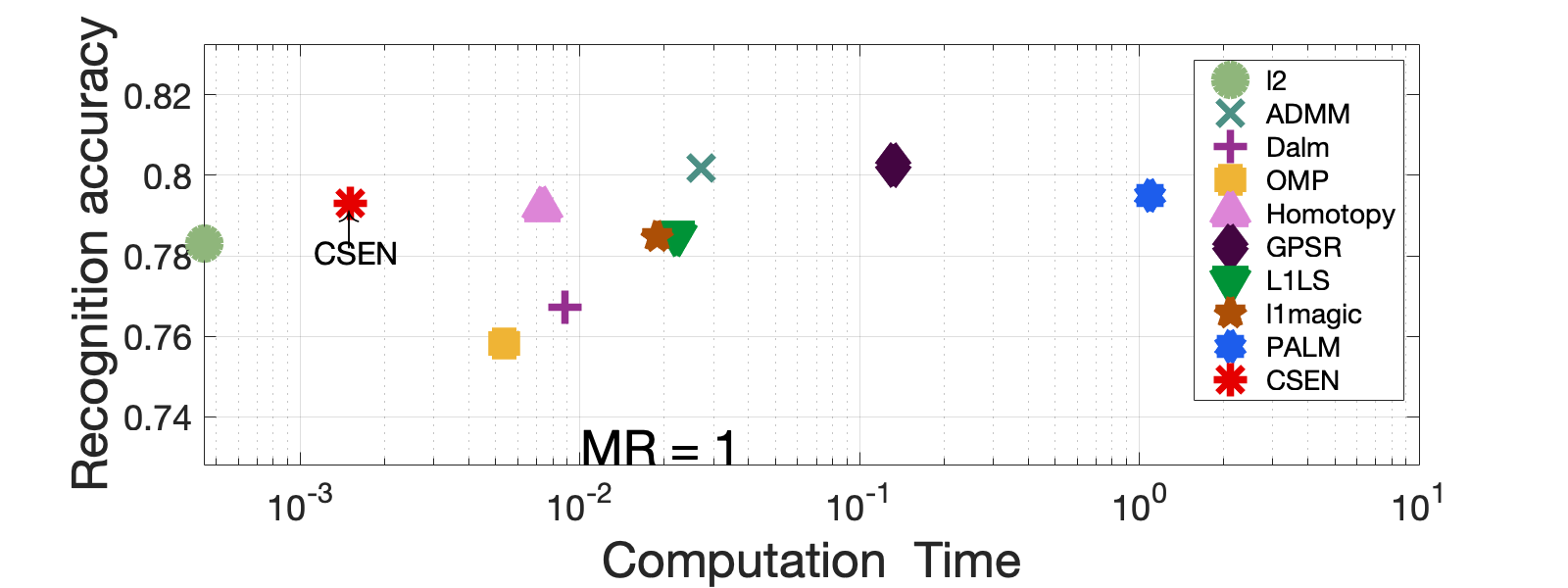}

  \includegraphics[width=0.95\linewidth]{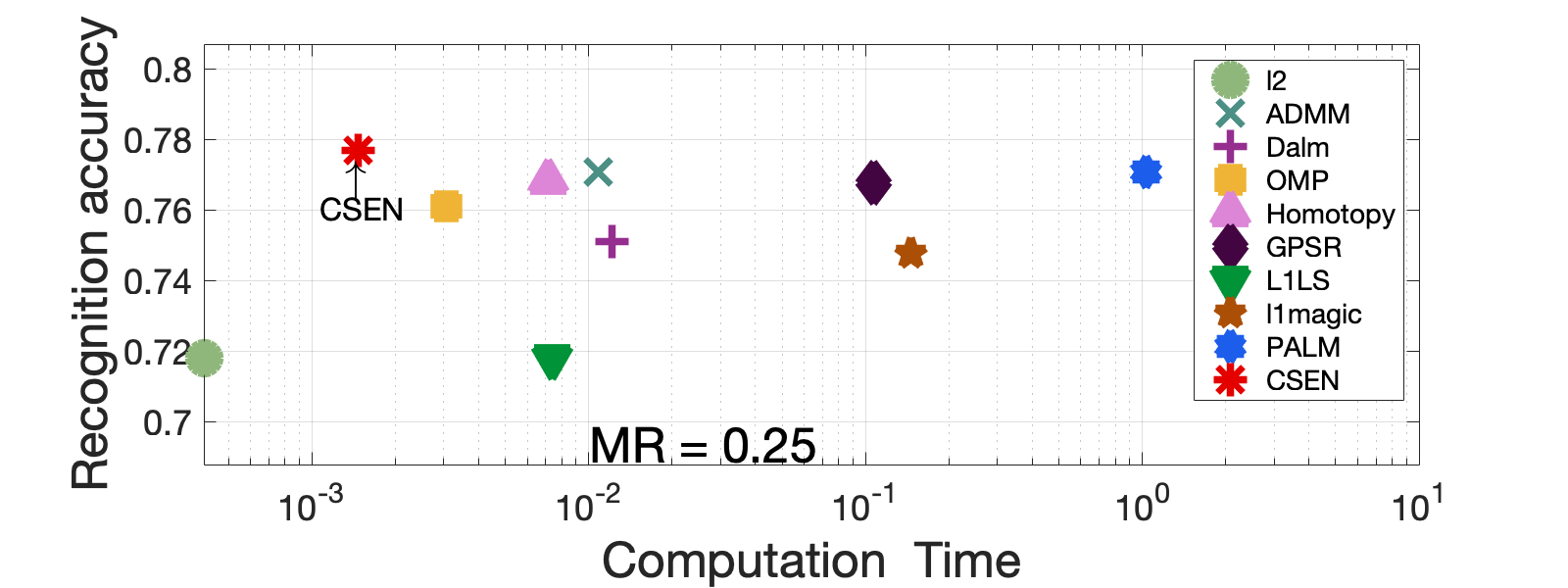}
  
  \includegraphics[width=0.95\linewidth]{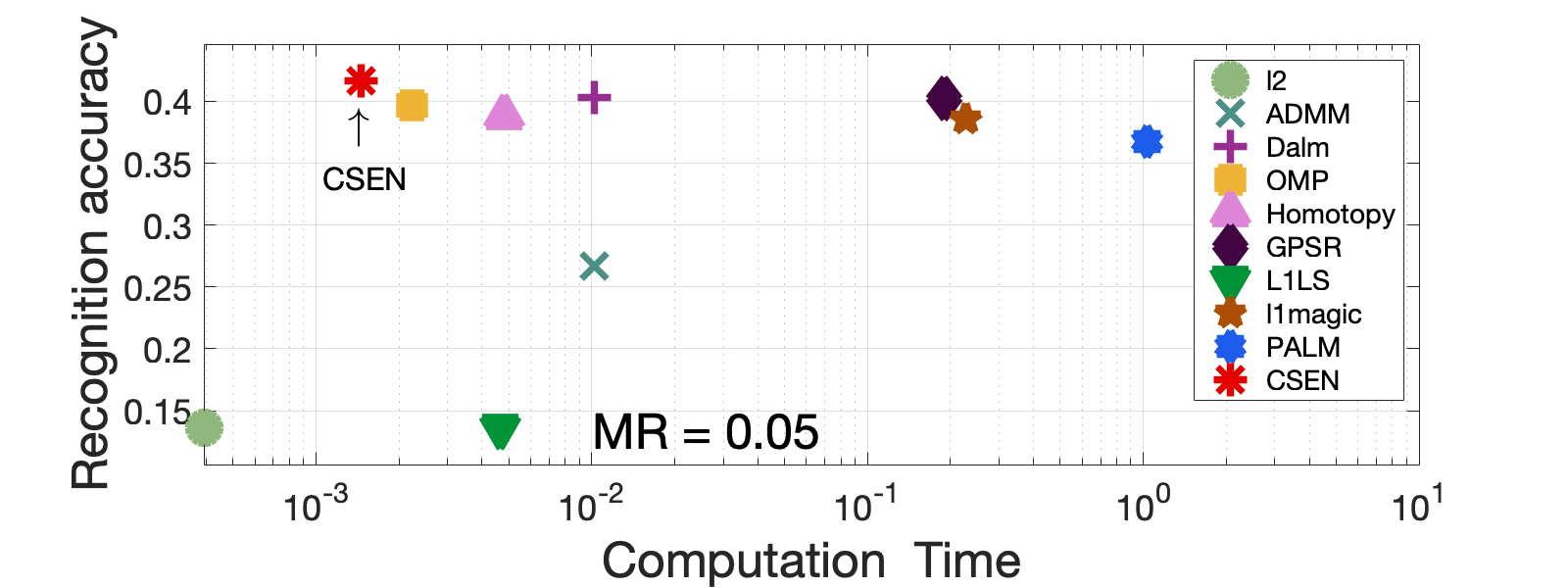}
  
  \caption{Reconstruction accuracy vs. process time comparison of Algorithms in CelebA database.}
\label{fig:CelebA}
\end{figure}

In dictionary-based classification designs, the samples of a specific class are stacked in the dictionary as atoms with pre-defined indices, e.g., the atoms belonging to a particular class can be located in concatenate manner. Consequently, in sparse representation based classification, instead of using $\ell_1$-minimization in Eq. (\ref{lasso}), group $\ell_1$-minimization can be introduced as follows,
\begin{equation}
    \min_\mathbf{x}  \left \{ \left \|  \mathbf{D}\mathbf{x}-\mathbf{y} \right \|_2^2 + \lambda \sum_{i=1}^{c}\left \|\mathbf{x_{Gi}} \right \|_2  \right \}.    
\end{equation}
where $\mathbf{x_{Gi}}$ is the group of coefficients corresponds to class $i$. Hence, the MSE cost function in Eq. (\ref{cost}) can be modified accordingly:
\begin{equation}
E(\mathbf{x}) =  \sum_p (\mathcal{P}_{\Theta}\left (\mathbf{x} \right )_p- v_p)^2 + \lambda \sum_{i=1}^{c}\left \|\mathcal{P}_{\Theta}\left (\mathbf{x} \right )_{Gi} \right \|_2. 
\end{equation}
To approximate this, a simple average pooling can be applied after the last layer of CSEN which is then followed by SoftMax function to produce class probabilities. Therefore, the modified cost function with Cross-Entropy loss at the output would be:
$E(\mathbf{x}) = -\sum_i^C t_ilog \left (\mathcal{P}_{\Theta}(\mathbf{x}) \right )$  where $t_i$ and $\mathcal{P}_{\Theta}(\mathbf{x})$ are the real and the predicted values by CSEN, respectively, for class $i \in C$. By this way, the modified network can directly produce the predicted class labels as the output.

In the experiments, we have used Yale-B \cite{yale} and CelebA \cite{CelebA} databases. In Yale-B dataset, there are $2414$ face images with 38 identities; and a sub-set of CelebA is chosen with 5600 images and 200 identities. The face recognition experiments are repeated 5 times with samples randomly selected to build the dictionary, train, and test sets with 32, 16, 16, and 8, 12, 8 samples each for Yale-B and CelebA, respectively, and 25\% of training data is separated as validation set. The selected sub-set of CelebA dataset is also different between each repeated run. For Yale-B database, we use vectorized images in the dictionary. Earlier studies reported that both SRC and CRC techniques achieve a high recognition accuracy such as 97 - 98\%, especially for high MR rate scenarios ($m/d > 0.25$ for $\mathbf{A} \in \mathbb{R}^{m \times d}$). On the other hand, for CalebA both CRC and SRC solution tends to fail when we use raw atoms in the dictionary without extracting descriptive features. This is why in this study, we propose to use a more representative dictionary. Instead of using raw images, the atoms consist of more descriptive features extracted by a neural network-based face feature extractor in the library~\cite{dlib}. The proposed method is compared against CRC and SRC techniques with the following 7 state-of-the-art SR solver: ADDM \cite{ADMM}, Dalm \cite{fast}, OMP \cite{fast}, Homotopy \cite{homotopy}, GPSR \cite{gpsr}, L1LS \cite{l1ls}, $\ell_1$-magic \cite{l1magic}, and Palm \cite{fast}.

Overall, when we perform experiments in two facial image databases, Yale-B and CalebA for different MRs, the CSEN based classification proves to be very stable; and in all MRs, it gives the highest or comparable recognition accuracy to the highest ones for all experiments as presented in Figure \ref{fig:Yale-B} and \ref{fig:CelebA}. Furthermore, it is significantly superior in terms of computational speed when compared with SRC solutions.

\begin{figure}[]
\centering
\includegraphics[width=0.65\linewidth]{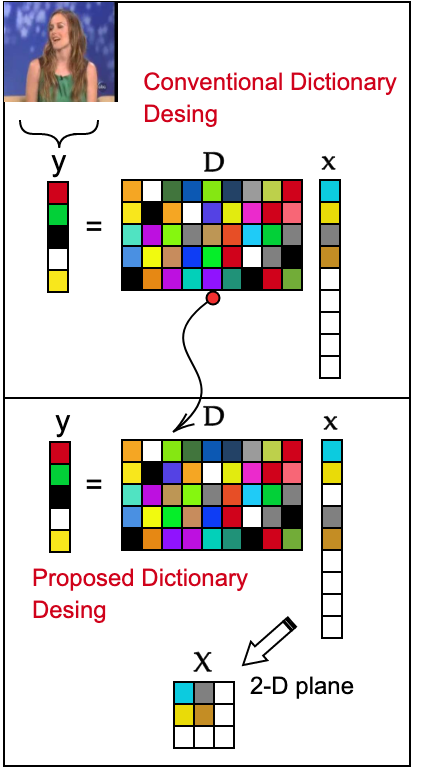}

  \caption{The graphical representation of proposed dictionary design vs. conventional design for face recognition problem.}
\label{fig:proposeddic}
\end{figure}

To be able to use the same CSEN designs introduced in Section \ref{Proposed}, we re-order the positions of the atoms, i.e., in the representative sparse codes corresponding non-zero coefficients remain next to each other in 2-D plane. A simplified illustration on the comparison of conventional dictionary design and the proposed design for sparse representation based classification is shown in Figure \ref{fig:proposeddic}. Defined sparse code sizes and their representations in 2-D grid for Yale-B and CelebA datasets are also given in Table \ref{facedata}. 
\begin{table}[]
\centering
\caption{Utilized face recognition benchmark datasets are given with their corresponding mask size and number of samples in dictionary, training, and testing per class.}
\begin{tabular}{@{}ccccc@{}}
Dataset & \begin{tabular}[c]{@{}c@{}}Dictionary\\ Samples\end{tabular} & Train Samples & Test Samples & SC Size in 2D-Plane \\ \hline \hline
Yale-B    & 32                                                           & 16            & 16           & 16 x 76   \\
CelebA  & 8                                                            & 12            & 8            & 8 x 200   \\
\end{tabular}
\label{facedata}
\end{table}
\subsection{Experiment III: Learning-aided Compressive Sensing}
\label{Learning-aided}

\begin{figure*}[]
 \centering
  \includegraphics[width=0.7\linewidth]{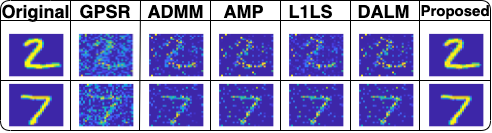}
  
 \caption{Examples from MNIST that are compressively sensed, and then reconstructed at MR=$0.25$.}
\label{fig:CSEN general examples}
\end{figure*}

\begin{figure}[h]
\centering
  \includegraphics[width=0.95\linewidth]{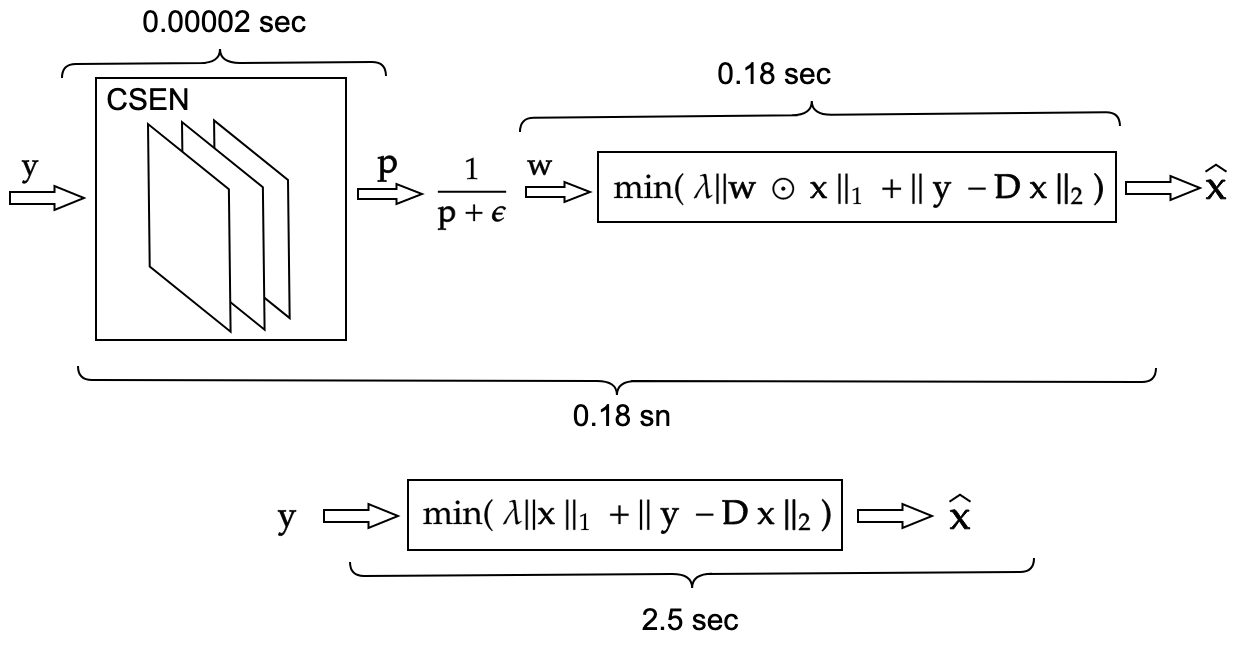}
  \caption{(Top) Proposed Compressive Sensing Reconstruction. (Bottom) Traditional $\ell_1$-minimization based CS-recovery.}
\label{fig:Networkaided-time}
\end{figure}

\begin{figure}[h]
\centering
  \includegraphics[width=0.9\linewidth]{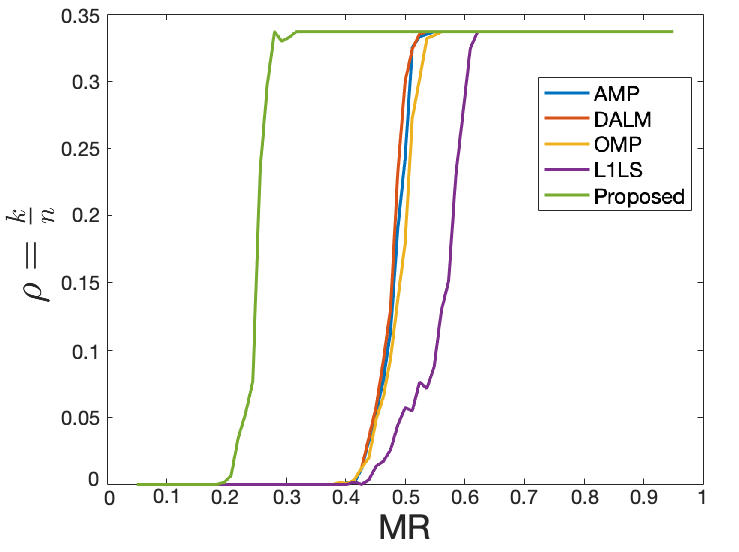}
 \caption{Phase Transition of the Algorithms. }
\label{fig:phase}
\end{figure}

As the experimental setup, we randomly choose, sparse signals, $\mathbf{x}$ in the MNIST database and use the Gaussian measurement matrix, $\mathbf{A}$ to simulate the CS, i.e., $\mathbf{y}=\mathbf{A} \mathbf{ x}$. Then, we recover the sparse signal from $\mathbf{y}$ by using the aforementioned state-of-the-art SR tools and the proposed weighted $\ell_1$-minimization Eq. \eqref{weighted-lasso}, where the weights $\mathbf{w}$ are obtained using CSEN output such that $\mathbf{w} = \frac{1}{\mathbf{p} + \epsilon}$. The two examples where signals are compressively sensed with $MR=0.25$ and their estimated versions by different SR methods are shown in Figure \ref{fig:CSEN general examples}. It is clear that the proposed approach recovers the sparse signal with the best quality while the other state-of-art SR techniques perform poorly. Figure \ref{fig:Networkaided-time} shows an illustration of how proposed compressive sensing reconstruction scheme differs from traditional compressive sensing recovery setup. Using the output of CSEN as prior information not only provide more accurate signal recovery, but also faster convergence of iterative sparse signal recovery such as $\ell_1$-minimization. 
Furthermore, we draw the estimated phase transition of the algorithms in Figure \ref{fig:phase} using an experimental setup whose procedure is explained in \cite{AMP}. Briefly summarizing the procedure, a grid of (MR, $\rho$) is generated for each algorithm, with 20 independent realization of the problem: according to their sparsity ratios, $\rho$, randomly chosen sparse signal $\mathbf{x}$, among $10000$ MNIST test images, are compressively sensed with independent realization of measurement matrices. Then, they are recovered using the competing algorithms, and each realization is considered a success for the specific algorithm if
\begin{equation}
\frac{\left \| \mathbf{x} - \mathbf{{\hat{x}}} \right \|_2}{\left \| \mathbf{x} \right \|} \leq \text{tol}
\end{equation}
where $\text{tol}$ is a predefined parameter, we choose $\text{tol}=10^{-1}$ in our experiments. For a specific algorithm, we draw the phase transition in the border where a $50\%$ success rate is achieved. The procedure is similar to \cite{AMP}, with the exception that they repeated the experiment only once, while we repeat it 100 times for each method, except L1LS due to its infeasibly high computational cost (it took almost two weeks with an ordinary computer). With an accurate SR algorithm, we expect the transition border to be close to the left-top corner in the phase transition graph because it is a good indicator that the algorithm performs well in low MRs and with high sparsity ratio, $\rho$. From the Figure, one can easily deduce that the proposed CS-reconstruction approach clearly outperforms all competing state-of-the-art SR reconstruction methods.
\section{Discussion}
\label{Discussion}

\subsection{Sparse Modelling vs Structural Sparse Modelling }

The first generation CS-recovery or sparse representation methods only use the information that the signal, we encounter in real life, is sparse in a proper domain or dictionary. These models do not utilize any further assumptions about the sparse signal, $\mathbf{x}$, in signal recovery or support estimation. Therefore, they only impose sparsity to the signal to have support set with elements in arbitrary location i.e., $\min \left \| \mathbf{x} \right \|_0 ~\text{s.t.} ~ \mathbf{Dx} = \mathbf{y}$. However, most sparse signals we face in practical applications exhibit a kind of structure. In second-generation sparse representation models, researchers realized that in addition to arbitrary sparsity any prior information about the sparse code can be used in modeling more advance recovery schemes \cite{structured_sparse, structured_sparse2}. For instance, the indices of the non-zero wavelet coefficients of an image mostly exhibit grouping effect \cite{group_sparse}. This kind of group sparsity patterns can be imposed by designing the optimization problem involving mixed norm minimization problems \cite{mixed_norm} instead of simple $\ell_1$-norm. On the other hand, more complex sparsity structures require more complex model design.

This work proposes an alternative solution to the hand-crafted model-based sparse signal recovery approaches, to be able to learn the pattern inside sparse code (or structural sparse signals), $\mathbf{x}$ by a Machine Learning technique. This proof of the concept work in which the performance is tested over 3 real datasets, MNIST, Yale and CelabA, validates the possibility of such learning and deserves further investigation in different sparse representation problems.

\subsection{Unrolling Deep Models vs CSEN}
The most common approaches to reconstruct sparse signals, $\mathbf{x}$ from the given measurements, $\mathbf{y}$ with a fixed dictionary $\mathbf{D}$ can be listed as follows: i) Convex Relaxation (or $\ell_1$ minimization) such as Basis Pursuit \cite{BP}:  $\min_{\mathbf{x}} \left \| \mathbf{x} \right \| ~\text{s.t.} ~    \mathbf{y}=\mathbf{Dx}$ or Basis Pursuit Denoising (BPDN) \cite{BP}: $\min_{\mathbf{x}} \left \| \mathbf{x} \right \| ~\text{s.t.} ~    \left \| \mathbf{y} - \mathbf{Dx} \right \|  \leq \epsilon$, where $\epsilon$ is a small  constant. ii) Greedy Algorithms such as matching pursuit  (MP) \cite{mallat1993}, Orthogonal Matching Pursuit (OMP) \cite{OMP}, Compressive Sampling Matched Pursuit  (CoSaMP) \cite{cosamp}. iii) Bayesian Framework \cite{bayesian}, etc. These conventional algorithms dealing with sparse inverse problems work in an iterative manner; for instance, most convex relaxation techniques such as BPDN,  minimize the data fidelity term (e.g., $\ell_2$-norm) and sparsifying term (e.g., $\ell_1$-norm) in an alternating manner in each iteration. Therefore, these schemes suffer from computational complexity and not suitable for real-time applications.

Along with the traditional approaches listed above, Deep Learning methods used in this domain have recently become very popular: $ \mathbf{ \hat{x} } \leftarrow    \mathcal{P} \left ( \mathbf{y} \right ) $, where $\mathcal{P}$ is a learned mapping from $m$ dimensional compressed domain to $n$ dimensional sparse domain. These techniques are built on the idea that the performance of existing convex relaxation can further be improved by reducing the number of iterations and enhancing the reconstruction accuracy. The key idea is that both the possible denoiser matrices, $\mathbf{ B }$ (responsible from dealing with data fidelity term) such as $\mathbf{D}^T$, or  $\left ( \mathbf{D}^T \mathbf{D} + \lambda \mathbf{I} \right )^{-1} \mathbf{D}^T$ where $\lambda$ is the regularization parameter, and the thresholding values (responsible from sparsifying) can be learned from the training data using a deep network generally with dense layers. For instance, the first example of this type is Learned-ISTA (LISTA) \cite{LISTA} which is built upon iterative soft-thresholding algorithm (ISTA) \cite{ISTA}. These category of methods, also called unrolled deep models, design networks in iterative manner are powerfull tools for sparse signal recovery.

However, in many practical applications, we may either no need to estimate the sparse signal itself or not have a large amount of training data for deep unrolling networks. In that manner, CSEN provides a third approach, by directly estimating the support set via a compact design which requires less computational power, memory and training set. It exhibits very good performance, especially in the problems that include sparse representation with sparse codes having structural patterns. The other advantage of the compact design with convolutional layers is that it more stable against noise compared to unrolled deep models which include dense layers.

\subsection{Proxy signal vs measurement vector as input to CSEN}
The proposed support estimation scheme utilizes proxy $\mathbf{ \tilde{x}} = \mathbf{By}$ as input to convolutional layers.  Making inference directly on proxy using ML approach has been recently reported by several studies. For example, the study in \cite{degerli,inference} proposed to perform reconstruction-free image classification on proxy, and the study in \cite{deepinCS} performed signal reconstruction using proxy as an input to a deep fully convolutional network. Furthermore, proxy $\mathbf{\Tilde{x}}$ can be learnt by fully-connected dense layers as presented in \cite{degerli}. However, this brings additional complexity and training the network may cause over-fitting with limited number of training data. As in \cite{degerli}, they had to adapt by first training the fully-connected layers or try to freeze the other layers during the training.

On the other hand, choosing the denoiser matrix, $\mathbf{B}$, is another design problem. For example, in \cite{degerli,inference} the authors use $\mathbf{B=D^T}$ as denoiser to obtain proxy. We reported the results in this paper for denoiser matrix,  $ \mathbf{B} =  \left ( \mathbf{D}^T \mathbf{D} + \lambda \mathbf{I} \right )^{-1} \mathbf{D}^T$, because it gives slightly more stable performance over $\mathbf{B}$. 

\section{Conclusions}
\label{conclusion}
Sparse support estimators that work based on traditional sparse signal recovery techniques suffer from computational complexity and noise. Moreover, they tend to fail at low MRs completely.  The proposed CSENs can be considered as reconstruction-free and non-iterative support estimators. 
Of course, despite their high computational complexity, recent state-of-the-art deep signal reconstruction algorithms may be a cure to sparse recovery methods. However, they are still redundant if SR is not the main concern. In addition, such deep networks often require a large amount of training data that is not available in many practical applications. To address these drawbacks and limitations, in this study, we introduce a novel learning-based support estimators, which have compact network designs. The highlights of the proposed system are as follows: i) Signal reconstruction-free support estimation where sparse estimation can be done in a feed-forward manner, non-iteratively at a low cost.  ii) Compact network designs enabling efficient learning even from a small-size training set. iii) The proposed solution is generic; it could be used in any support estimation task such as SE based classification.

\ifCLASSOPTIONcaptionsoff
  \newpage
\fi



%
{\small
\bibliographystyle{IEEEtran}
\bibliography{egbib}

\begin{thebibliography}{10}
\providecommand{\url}[1]{#1}
\csname url@samestyle\endcsname
\providecommand{\newblock}{\relax}
\providecommand{\bibinfo}[2]{#2}
\providecommand{\BIBentrySTDinterwordspacing}{\spaceskip=0pt\relax}
\providecommand{\BIBentryALTinterwordstretchfactor}{4}
\providecommand{\BIBentryALTinterwordspacing}{\spaceskip=\fontdimen2\font plus
\BIBentryALTinterwordstretchfactor\fontdimen3\font minus
  \fontdimen4\font\relax}
\providecommand{\BIBforeignlanguage}[2]{{%
\expandafter\ifx\csname l@#1\endcsname\relax
\typeout{** WARNING: IEEEtran.bst: No hyphenation pattern has been}%
\typeout{** loaded for the language `#1'. Using the pattern for}%
\typeout{** the default language instead.}%
\else
\language=\csname l@#1\endcsname
\fi
#2}}
\providecommand{\BIBdecl}{\relax}
\BIBdecl

\bibitem{CS1}
D.~L. Donoho \emph{et~al.}, ``Compressed sensing,'' \emph{IEEE Transactions on
  information theory}, vol.~52, no.~4, pp. 1289--1306, 2006.

\bibitem{CS2}
E.~J. Cand{\`e}s \emph{et~al.}, ``Compressive sampling,'' in \emph{Proceedings
  of the International Congress of Mathematicians}, vol.~3, 2006, pp.
  1433--1452.

\bibitem{equivalent}
G.~Li, Z.~Zhu, D.~Yang, L.~Chang, and H.~Bai, ``On projection matrix
  optimization for compressive sensing systems,'' \emph{IEEE Transactions on
  Signal Processing}, vol.~61, no.~11, pp. 2887--2898, 2013.

\bibitem{Elad}
M.~Elad, \emph{Sparse and redundant representations: from theory to
  applications in signal and image processing}.\hskip 1em plus 0.5em minus
  0.4em\relax Springer Science \& Business Media, 2010.

\bibitem{SE1}
W.~Wang, M.~J. Wainwright, and K.~Ramchandran, ``Information-theoretic limits
  on sparse support recovery: Dense versus sparse measurements,'' in \emph{2008
  IEEE International Symposium on Information Theory}.\hskip 1em plus 0.5em
  minus 0.4em\relax IEEE, 2008, pp. 2197--2201.

\bibitem{SE2}
J.~Haupt and R.~Baraniuk, ``Robust support recovery using sparse compressive
  sensing matrices,'' in \emph{2011 45th Annual Conference on Information
  Sciences and Systems}.\hskip 1em plus 0.5em minus 0.4em\relax IEEE, 2011, pp.
  1--6.

\bibitem{SE3}
G.~Reeves and M.~Gastpar, ``Sampling bounds for sparse support recovery in the
  presence of noise,'' in \emph{2008 IEEE International Symposium on
  Information Theory}.\hskip 1em plus 0.5em minus 0.4em\relax IEEE, 2008, pp.
  2187--2191.

\bibitem{OMP}
J.~A. Tropp and A.~C. Gilbert, ``Signal recovery from random measurements via
  orthogonal matching pursuit,'' \emph{IEEE Transactions on information
  theory}, vol.~53, no.~12, pp. 4655--4666, 2007.

\bibitem{cosamp}
D.~Needell and J.~A. Tropp, ``Cosamp: Iterative signal recovery from incomplete
  and inaccurate samples,'' \emph{Applied and computational harmonic analysis},
  vol.~26, no.~3, pp. 301--321, 2009.

\bibitem{Volkan}
J.~Scarlett and V.~Cevher, ``Limits on support recovery with probabilistic
  models: An information-theoretic framework,'' \emph{IEEE Transactions on
  Information Theory}, vol.~63, no.~1, pp. 593--620, 2016.

\bibitem{SRC2}
J.~Wright, Y.~Ma, J.~Mairal, G.~Sapiro, T.~S. Huang, and S.~Yan, ``Sparse
  representation for computer vision and pattern recognition,''
  \emph{Proceedings of the IEEE}, vol.~98, no.~6, pp. 1031--1044, 2010.

\bibitem{SRC1}
J.~Wright, A.~Y. Yang, A.~Ganesh, S.~S. Sastry, and Y.~Ma, ``Robust face
  recognition via sparse representation,'' \emph{IEEE transactions on pattern
  analysis and machine intelligence}, vol.~31, no.~2, pp. 210--227, 2008.

\bibitem{SS1}
B.~Khalfi, B.~Hamdaoui, M.~Guizani, and N.~Zorba, ``Efficient spectrum
  availability information recovery for wideband dsa networks: A weighted
  compressive sampling approach,'' \emph{IEEE Transactions on Wireless
  Communications}, vol.~17, no.~4, pp. 2162--2172, 2018.

\bibitem{SS2}
B.~Hamdaoui, B.~Khalfi, and M.~Guizani, ``Compressed wideband spectrum sensing:
  Concept, challenges, and enablers,'' \emph{IEEE Communications Magazine},
  vol.~56, no.~4, pp. 136--141, 2018.

\bibitem{radar1}
A.~C. Gurbuz, J.~H. McClellan, and W.~R. Scott, ``A compressive sensing data
  acquisition and imaging method for stepped frequency {GPR}s,'' \emph{IEEE
  Transactions on Signal Processing}, vol.~57, no.~7, pp. 2640--2650, 2009.

\bibitem{fully}
J.~Long, E.~Shelhamer, and T.~Darrell, ``Fully convolutional networks for
  semantic segmentation,'' in \emph{Proceedings of the IEEE conference on
  computer vision and pattern recognition}, 2015, pp. 3431--3440.

\bibitem{reconnet}
K.~Kulkarni, S.~Lohit, P.~Turaga, R.~Kerviche, and A.~Ashok, ``Reconnet:
  Non-iterative reconstruction of images from compressively sensed
  measurements,'' in \emph{Proceedings of the IEEE Conference on Computer
  Vision and Pattern Recognition}, 2016, pp. 449--458.

\bibitem{Lamp}
M.~Borgerding, P.~Schniter, and S.~Rangan, ``Amp-inspired deep networks for
  sparse linear inverse problems,'' \emph{IEEE Transactions on Signal
  Processing}, vol.~65, no.~16, pp. 4293--4308, 2017.

\bibitem{AMP}
D.~L. Donoho, A.~Maleki, and A.~Montanari, ``Message-passing algorithms for
  compressed sensing,'' \emph{Proceedings of the National Academy of Sciences},
  vol. 106, no.~45, pp. 18\,914--18\,919, 2009.

\bibitem{yale}
A.~S. Georghiades, P.~N. Belhumeur, and D.~J. Kriegman, ``From few to many:
  Illumination cone models for face recognition under variable lighting and
  pose,'' \emph{IEEE Transactions on Pattern Analysis \& Machine Intelligence},
  no.~6, pp. 643--660, 2001.

\bibitem{CelebA}
Z.~Liu, P.~Luo, X.~Wang, and X.~Tang, ``Deep learning face attributes in the
  wild,'' in \emph{Proceedings of International Conference on Computer Vision
  (ICCV)}, December 2015.

\bibitem{collaborative}
L.~Zhang, M.~Yang, and X.~Feng, ``Sparse representation or collaborative
  representation: Which helps face recognition?'' in \emph{2011 International
  conference on computer vision}.\hskip 1em plus 0.5em minus 0.4em\relax IEEE,
  2011, pp. 471--478.

\bibitem{priori1}
O.~D. Escoda, L.~Granai, and P.~Vandergheynst, ``On the use of a priori
  information for sparse signal approximations,'' \emph{IEEE transactions on
  signal processing}, vol.~54, no.~9, pp. 3468--3482, 2006.

\bibitem{recursive1}
N.~Vaswani and J.~Zhan, ``Recursive recovery of sparse signal sequences from
  compressive measurements: A review,'' \emph{IEEE Transactions on Signal
  Processing}, vol.~64, no.~13, pp. 3523--3549, 2016.

\bibitem{spark}
D.~L. Donoho and M.~Elad, ``Optimally sparse representation in general
  (nonorthogonal) dictionaries via $ell_1$ minimization,'' \emph{Proceedings of
  the National Academy of Sciences}, vol. 100, no.~5, pp. 2197--2202, 2003.

\bibitem{BP}
S.~S. Chen, D.~L. Donoho, and M.~A. Saunders, ``Atomic decomposition by basis
  pursuit,'' \emph{SIAM review}, vol.~43, no.~1, pp. 129--159, 2001.

\bibitem{candesRIP}
E.~J. Candes, ``The restricted isometry property and its implications for
  compressed sensing,'' \emph{Comptes rendus mathematique}, vol. 346, no. 9-10,
  pp. 589--592, 2008.

\bibitem{BPDN}
S.~S. Chen, D.~L. Donoho, and M.~A. Saunders, ``Atomic decomposition by basis
  pursuit,'' \emph{SIAM review}, vol.~43, no.~1, pp. 129--159, 2001.

\bibitem{dantzig}
E.~Candes, T.~Tao \emph{et~al.}, ``The dantzig selector: Statistical estimation
  when p is much larger than n,'' \emph{The annals of Statistics}, vol.~35,
  no.~6, pp. 2313--2351, 2007.

\bibitem{candesRIP2}
E.~J. Candes, J.~K. Romberg, and T.~Tao, ``Stable signal recovery from
  incomplete and inaccurate measurements,'' \emph{Communications on Pure and
  Applied Mathematics: A Journal Issued by the Courant Institute of
  Mathematical Sciences}, vol.~59, no.~8, pp. 1207--1223, 2006.

\bibitem{Eldar}
Y.~C. Eldar and G.~Kutyniok, \emph{Compressed sensing: theory and
  applications}.\hskip 1em plus 0.5em minus 0.4em\relax Cambridge University
  Press, 2012.

\bibitem{dantzig2}
M.~S. Asif and J.~Romberg, ``On the lasso and dantzig selector equivalence,''
  in \emph{2010 44th Annual Conference on Information Sciences and Systems
  (CISS)}.\hskip 1em plus 0.5em minus 0.4em\relax IEEE, 2010, pp. 1--6.

\bibitem{lasso}
R.~Tibshirani, ``Regression shrinkage and selection via the lasso,''
  \emph{Journal of the Royal Statistical Society: Series B (Methodological)},
  vol.~58, no.~1, pp. 267--288, 1996.

\bibitem{lasso-stable}
E.~J. Candes and Y.~Plan, ``A probabilistic and ripless theory of compressed
  sensing,'' \emph{IEEE transactions on information theory}, vol.~57, no.~11,
  pp. 7235--7254, 2011.

\bibitem{compressedanomaly}
N.~Durgin, R.~Grotheer, C.~Huang, S.~Li, A.~Ma, D.~Needell, and J.~Qin,
  ``Compressed anomaly detection with multiple mixed observations,'' in
  \emph{Research in Data Science}.\hskip 1em plus 0.5em minus 0.4em\relax
  Springer, 2019, pp. 211--237.

\bibitem{yamacmalicious}
M.~Yama{\c{c}}, B.~Sankur, and A.~T. Cemgil, ``Malicious users discrimination
  in organized attacks using structured sparsity,'' in \emph{2017 25th European
  Signal Processing Conference (EUSIPCO)}.\hskip 1em plus 0.5em minus
  0.4em\relax IEEE, 2017, pp. 266--270.

\bibitem{CDMA}
B.~Shim and B.~Song, ``Multiuser detection via compressive sensing,''
  \emph{IEEE Communications Letters}, vol.~16, no.~7, pp. 972--974, 2012.

\bibitem{Noma1}
O.~O. Oyerinde, ``Multiuser detector for uplink grant free noma systems based
  on modified subspace pursuit algorithm,'' in \emph{2018 12th International
  Conference on Signal Processing and Communication Systems (ICSPCS)}.\hskip
  1em plus 0.5em minus 0.4em\relax IEEE, 2018, pp. 1--6.

\bibitem{Noma2}
B.~Wang, L.~Dai, Y.~Zhang, T.~Mir, and J.~Li, ``Dynamic compressive
  sensing-based multi-user detection for uplink grant-free noma,'' \emph{IEEE
  Communications Letters}, vol.~20, no.~11, pp. 2320--2323, 2016.

\bibitem{radar2}
M.~Yama{\c{c}}, M.~Orhan, B.~Sankur, A.~S. Turk, and M.~Gabbouj, ``Through the
  wall target detection/monitoring from compressively sensed signals via
  structural sparsity,'' in \emph{5th International Workshop on Compressed
  Sensing applied to Radar, Multimodal Sensing,and Imaging}, 2018.

\bibitem{exact1}
K.~R. Rad, ``Nearly sharp sufficient conditions on exact sparsity pattern
  recovery,'' \emph{IEEE Transactions on Information Theory}, vol.~57, no.~7,
  pp. 4672--4679, 2011.

\bibitem{exact4}
M.~Wainwright, ``Information-theoretic bounds on sparsity recovery in the
  high-dimensional and noisy setting,'' in \emph{2007 IEEE International
  Symposium on Information Theory}.\hskip 1em plus 0.5em minus 0.4em\relax
  IEEE, 2007, pp. 961--965.

\bibitem{partial2}
G.~Reeves and M.~C. Gastpar, ``Approximate sparsity pattern recovery:
  Information-theoretic lower bounds,'' \emph{IEEE Transactions on Information
  Theory}, vol.~59, no.~6, pp. 3451--3465, 2013.

\bibitem{candes2009near}
E.~J. Cand{\`e}s, Y.~Plan \emph{et~al.}, ``Near-ideal model selection by
  $ell_1$ minimization,'' \emph{The Annals of Statistics}, vol.~37, no.~5A, pp.
  2145--2177, 2009.

\bibitem{OMP-exact}
T.~T. Cai and L.~Wang, ``Orthogonal matching pursuit for sparse signal recovery
  with noise.''\hskip 1em plus 0.5em minus 0.4em\relax Institute of Electrical
  and Electronics Engineers, 2011.

\bibitem{AMP-partial}
G.~Reeves and M.~Gastpar, ``The sampling rate-distortion tradeoff for sparsity
  pattern recovery in compressed sensing,'' \emph{IEEE Transactions on
  Information Theory}, vol.~58, no.~5, pp. 3065--3092, 2012.

\bibitem{MaximumCorrelation}
A.~K. Fletcher, S.~Rangan, and V.~K. Goyal, ``Necessary and sufficient
  conditions for sparsity pattern recovery,'' \emph{IEEE Transactions on
  Information Theory}, vol.~55, no.~12, pp. 5758--5772, 2009.

\bibitem{human-action}
T.~Guha and R.~K. Ward, ``Learning sparse representations for human action
  recognition,'' \emph{IEEE Transactions on Pattern Analysis and Machine
  Intelligence}, vol.~34, no.~8, pp. 1576--1588, 2011.

\bibitem{hyperspecral}
W.~Li and Q.~Du, ``A survey on representation-based classification and
  detection in hyperspectral remote sensing imagery,'' \emph{Pattern
  Recognition Letters}, vol.~83, pp. 115--123, 2016.

\bibitem{Modified-CS}
N.~Vaswani and W.~Lu, ``Modified-cs: Modifying compressive sensing for problems
  with partially known support,'' \emph{IEEE Transactions on Signal
  Processing}, vol.~58, no.~9, pp. 4595--4607, 2010.

\bibitem{coherence}
E.~Candes and J.~Romberg, ``Sparsity and incoherence in compressive sampling,''
  \emph{Inverse problems}, vol.~23, no.~3, p. 969, 2007.

\bibitem{phasetransition}
D.~Donoho and J.~Tanner, ``Observed universality of phase transitions in
  high-dimensional geometry, with implications for modern data analysis and
  signal processing,'' \emph{Philosophical Transactions of the Royal Society A:
  Mathematical, Physical and Engineering Sciences}, vol. 367, no. 1906, pp.
  4273--4293, 2009.

\bibitem{LISTA}
K.~Gregor and Y.~LeCun, ``Learning fast approximations of sparse coding,'' in
  \emph{Proceedings of the 27th International Conference on International
  Conference on Machine Learning}.\hskip 1em plus 0.5em minus 0.4em\relax
  Omnipress, 2010, pp. 399--406.

\bibitem{fast}
A.~Y. Yang, Z.~Zhou, A.~G. Balasubramanian, S.~S. Sastry, and Y.~Ma, ``Fast
  $ell_1$-minimization algorithms for robust face recognition,'' \emph{IEEE
  Transactions on Image Processing}, vol.~22, no.~8, pp. 3234--3246, 2013.

\bibitem{abadi2016tensorflow}
M.~Abadi, A.~Agarwal, P.~Barham, E.~Brevdo, Z.~Chen, C.~Citro, G.~S. Corrado,
  A.~Davis, J.~Dean, M.~Devin \emph{et~al.}, ``Tensorflow: Large-scale machine
  learning on heterogeneous distributed systems,'' \emph{arXiv preprint
  arXiv:1603.04467}, 2016.

\bibitem{kingma2014adam}
D.~P. Kingma and J.~Ba, ``Adam: A method for stochastic optimization,''
  \emph{arXiv preprint arXiv:1412.6980}, 2014.

\bibitem{dlib}
D.~E. King, ``Dlib-ml: A machine learning toolkit,'' \emph{Journal of Machine
  Learning Research}, vol.~10, no. Jul, pp. 1755--1758, 2009.

\bibitem{ADMM}
S.~Boyd, N.~Parikh, E.~Chu, B.~Peleato, J.~Eckstein \emph{et~al.},
  ``Distributed optimization and statistical learning via the alternating
  direction method of multipliers,'' \emph{Foundations and
  Trends{\textregistered} in Machine learning}, vol.~3, no.~1, pp. 1--122,
  2011.

\bibitem{homotopy}
D.~M. Malioutov, M.~Cetin, and A.~S. Willsky, ``Homotopy continuation for
  sparse signal representation,'' in \emph{Proceedings.(ICASSP'05). IEEE
  International Conference on Acoustics, Speech, and Signal Processing, 2005.},
  vol.~5.\hskip 1em plus 0.5em minus 0.4em\relax IEEE, 2005, pp. v--733.

\bibitem{gpsr}
M.~A. Figueiredo, R.~D. Nowak, and S.~J. Wright, ``Gradient projection for
  sparse reconstruction: Application to compressed sensing and other inverse
  problems,'' \emph{IEEE Journal of selected topics in signal processing},
  vol.~1, no.~4, pp. 586--597, 2007.

\bibitem{l1ls}
K.~Koh, S.-J. Kim, and S.~Boyd, ``An interior-point method for large-scale
  l1-regularized logistic regression,'' \emph{Journal of Machine learning
  research}, vol.~8, no. Jul, pp. 1519--1555, 2007.

\bibitem{l1magic}
E.~Candes and J.~Romberg, ``l1-magic: Recovery of sparse signals via convex
  programming,'' \emph{URL: www. acm. caltech. edu/l1magic/downloads/l1magic.
  pdf}, vol.~4, p.~14, 2005.

\bibitem{structured_sparse}
G.~Yu, G.~Sapiro, and S.~Mallat, ``Solving inverse problems with piecewise
  linear estimators: From gaussian mixture models to structured sparsity,''
  \emph{IEEE Transactions on Image Processing}, vol.~21, no.~5, pp. 2481--2499,
  2012.

\bibitem{structured_sparse2}
R.~G. Baraniuk, V.~Cevher, M.~F. Duarte, and C.~Hegde, ``Model-based
  compressive sensing,'' \emph{IEEE Transactions on Information Theory},
  vol.~56, no.~4, pp. 1982--2001, 2010.

\bibitem{group_sparse}
D.~Donoho and G.~Kutyniok, ``Microlocal analysis of the geometric separation
  problem,'' \emph{Communications on Pure and Applied Mathematics}, vol.~66,
  no.~1, pp. 1--47, 2013.

\bibitem{mixed_norm}
M.~Kowalski and B.~Torr{\'e}sani, ``Structured sparsity: from mixed norms to
  structured shrinkage,'' in \emph{SPARS'09-Signal Processing with Adaptive
  Sparse Structured Representations}, 2009.

\bibitem{mallat1993}
S.~G. Mallat and Z.~Zhang, ``Matching pursuits with time-frequency
  dictionaries,'' \emph{IEEE Transactions on signal processing}, vol.~41,
  no.~12, pp. 3397--3415, 1993.

\bibitem{bayesian}
S.~Ji, Y.~Xue, L.~Carin \emph{et~al.}, ``Bayesian compressive sensing,''
  \emph{IEEE Transactions on signal processing}, vol.~56, no.~6, p. 2346, 2008.

\bibitem{ISTA}
A.~Chambolle, R.~A. De~Vore, N.-Y. Lee, and B.~J. Lucier, ``Nonlinear wavelet
  image processing: variational problems, compression, and noise removal
  through wavelet shrinkage,'' \emph{IEEE Transactions on Image Processing},
  vol.~7, no.~3, pp. 319--335, 1998.

\bibitem{degerli}
A.~De{\u{g}}erli, S.~Aslan, M.~Yamac, B.~Sankur, and M.~Gabbouj,
  ``Compressively sensed image recognition,'' in \emph{2018 7th European
  Workshop on Visual Information Processing (EUVIP)}.\hskip 1em plus 0.5em
  minus 0.4em\relax IEEE, 2018, pp. 1--6.

\bibitem{inference}
S.~{Lohit}, K.~{Kulkarni}, and P.~{Turaga}, ``Direct inference on compressive
  measurements using convolutional neural networks,'' in \emph{2016 IEEE
  International Conference on Image Processing (ICIP)}, Sep. 2016, pp.
  1913--1917.

\bibitem{deepinCS}
A.~Mousavi and R.~G. Baraniuk, ``Learning to invert: Signal recovery via deep
  convolutional networks,'' in \emph{2017 IEEE International Conference on
  Acoustics, Speech and Signal Processing (ICASSP)}.\hskip 1em plus 0.5em minus
  0.4em\relax IEEE, 2017, pp. 2272--2276.

\end{thebibliography}
}


%








\end{document}